\documentclass{aims} 
\usepackage{amsmath}
\usepackage{paralist}
\usepackage[misc]{ifsym}
\usepackage{epsfig} 
\usepackage{epstopdf} 

\usepackage{graphicx}
\usepackage{setspace}
\usepackage{lineno}
\usepackage{enumerate}
\usepackage{enumitem}
\usepackage{placeins}
\usepackage{amsfonts}
\usepackage{tikz}
\usepackage{multirow}   
\usepackage{makecell}   
\usepackage{array} 
\usepackage{rotating}
\usepackage{xcolor}
\usepackage{caption}
\usepackage{empheq}

\usepackage[colorlinks=true]{hyperref}
\hypersetup{urlcolor=blue, citecolor=red}
\allowdisplaybreaks

\def\currentvolume{X}
 \def\currentissue{X}
  \def\currentyear{200X}
   \def\currentmonth{XX}
    \def\ppages{X-XX}
     \def\DOI{10.3934/xx.xxxxxxx}

\theoremstyle{definition}

\title[DIP for PAT can mitigate limited-view artifacts]
{Deep Image Prior for photoacoustic tomography can mitigate limited-view artifacts} 

\author[H. Pulkkinen, J. Poimala, L. Kunyanksy, J. Gröhl and A. Hauptmann]{}


\thanks{$^*$Corresponding author: Hanna Pulkkinen}

\begin{document}
\maketitle

\centerline{\scshape
Hanna Pulkkinen$^{{\href{mailto:hanna.j.pulkkinen@oulu.fi}{\textrm{\Letter}}}*1}$
Jenni Poimala$^{{\href{mailto:jenni.poimala@uef.fi}{\textrm{\Letter}}}2}$
Leonid Kunyansky$^{{\href{mailto:leonid@arizona.edu}{\textrm{\Letter}}}3}$}
\centerline{\scshape
Janek Gröhl$^{{\href{mailto:j.groehl@eni-g.de}{\textrm{\Letter}}}4}$
and Andreas Hauptmann$^{{\href{mailto:andreas.hauptmann@oulu.fi}{\textrm{\Letter}}}1,5}$}

\medskip

{\footnotesize
 \centerline{$^1$Research Unit of Mathematical Sciences, University of Oulu, Finland}
} 

\medskip

{\footnotesize
 \centerline{$^2$Department of Technical Physics, University of Eastern Finland, Finland}
}

\medskip
{\footnotesize
\centerline{$^3$Department of Mathematics, University of Arizona, USA}}

\medskip
{\footnotesize
\centerline{$^4$ENI-G, a Joint Initiative of the University Medical Center Göttingen}}
{\footnotesize 
\centerline{ and the Max Planck Institute for Multidisciplinary Sciences,  Germany}

\medskip
{\footnotesize
\centerline{$^5$Department of Computer Science, University College London, U.K}}

\bigskip

 \centerline{(Communicated by Handling Editor)}


\begin{abstract}
We study the deep image prior (DIP) framework applied to photoacoustic tomography (PAT) as an unsupervised reconstruction approach to mitigate limited-view artifacts and noise commonly encountered in experimental settings. Efficient implementation is achieved by employing recently published fast forward and adjoint algorithms for circular measurement geometries. Initialization via a fast inverse and total variation (TV) regularization are applied to further suppress noise and mitigate overfitting. For comparison, we compute a classical TV reconstruction. Our experiments comprise simulated PAT measurements under limited-view geometries and varying levels of added noise as well as experimental measurements together with using a digital twin for quality assessment. Our findings suggest that DIP framework provides an effective unsupervised strategy for  robust PAT reconstruction even in the challenging case of a limited view geometry providing improvement in several quantitative measures over total variation reconstructions. 
\end{abstract}


\section{Introduction}

Photoacoustic tomography (PAT) is a hybrid imaging modality that combines pulsed laser illumination and ultrasound detection, enabling high-resolution imaging of soft biological tissue on a sub-millimeter scale. In PAT, short near-infrared light pulses are absorbed by chromophoric molecules such as melanin or hemoglobin, inducing thermoelastic expansion and generating acoustic pressure waves that are detected by an array of ultrasonic transducers.
Reconstructing an image of the initial pressure distribution from these acoustic measurements corresponds to solving the \textit{acoustic inverse source problem} of PAT~\cite{beard2011biomedical,wanglihong2021recent,hauptmann2024model}. This is described by the following initial value problem for the wave equation
\begin{align}
\left\{
\begin{alignedat}{2}
\left(\frac{\partial^2 }{\partial t^2} - c^2 \Delta\right) p(x,t) &{}= 0 
&\quad& \text{in } \Omega \times (0, T], \\
p(x,0) &{}= f(x) &\quad& \text{in } \Omega,  \\ 
\frac{\partial}{\partial t}p(x,0) &{}= 0 &\quad& \text{in } \Omega,
\end{alignedat}
\right.
\label{eqn:waveEq}
\end{align}
where $p(x,t)$ is the acoustic pressure at position $x$ and time $t$, $f(x)$ denotes the initial pressure distribution to be recovered, $c$ is constant speed of sound, $\Omega\subset\mathbb{R}^2$ is the computational domain with boundary $\Gamma=\partial\Omega$. The measurement is given by the measured time series on the boundary  $\Gamma$ as 
\begin{equation}
g(x,t) = p(x,t), \quad \text{where} (x,t) \in  \Gamma \times (0,T],    
\label{eqn:measurements}
\end{equation}
where $T$ is the total observation time, defining the time interval $(0,T]$. $x \in \Gamma$ represents detector location, and $g(x,t)$ denotes the measured acoustic data. 
While this problem is linear and well-posed under ideal imaging conditions, practical measurement setups often suffer from limited detector coverage, that means we can measure $g$ only a subset $\Gamma_m\subset\Gamma$. Additional detector subsampling and noise in the measurements make the inverse problem ill-posed and lead to artifacts such as streaking and blurring in reconstructed images~\cite{dreier2019operator, guan2020limited, rietberg2025artifacts}. 

The reconstruction quality in PAT is particularly sensitive to the number of detectors and their angular distributions. In many experimental settings, the sensor geometry is restricted by physical or anatomical constraints, leading to limited-view configurations and degraded image quality~\cite{dreier2019operator, guan2020limited}. 

Deep convolutional neural networks (CNNs) have achieved state-of-the-art performance in PAT image reconstruction tasks~\cite{antholzer2019deep,grohl2021deep, hauptmann2020deep,wanglihong2021recent}.
However, their success typically depends on supervised training, which requires access to large paired datasets of measurements and corresponding ground truth images. Such paired data are rarely available in PAT as obtaining ground truth data is inherently difficult due to the fact that true optimal absorption distribution cannot be measured \textit{in vivo}~\cite{grohl2023moving, grohl2025digital}. Photoacoustic data acquisition is also time-consuming, further limiting the availability of comprehensive paired training datasets. Consequently, many proposed solutions have been trained on gold-standard full-view reconstructions as reference, rather than a known ground-truth \cite{hauptmann2018model}. While these reference images enable training, they might transfer reconstruction bias and artifacts of the chosen algorithm to the learned model and hence limit the generalizability across different detector configurations, number of detectors, or noise characteristics. 

These inevitable shortcomings indicate that conventional supervised and learned iterative methods are not optimal in every scenario, which motivates the use of self- and unsupervised approaches such as the Deep Image Prior (DIP)~\cite{baguer2020computed,barbano2022educated,  dittmer2020regularization, ulyanov2018deep}. Unlike conventional supervised CNN methods that optimize the network parameters through extensive training with labeled data, DIP optimizes the parameters of an untrained convolutional neural network directly for a single measurement, using the CNN's architecture itself as an implicit prior. While originally proposed for denoising and natural image restoration as in Ulyanov et al.~\cite{ulyanov2018deep}, DIP has since been extended to various inverse problems in imaging \cite{baguer2020computed, barbano2022educated, dittmer2020regularization,lan2021compressed} and provides a completely unsupervised framework for imaging modalities such as PAT, where ground truth data is scarce. 

DIP-based approaches have been explored earlier in photoacoustic imaging. For example, DIP has been applied to photoacoustic microscopy (PAM) for image-domain enhancement of undersampled reconstructions, where it acts as a post-processing~\cite{vu2021deep}. For PAT, specifically with a focus on compressed sensing under sparse sampling conditions, DIP has been investigated using a matrix representation for the forward operator~\cite{lan2021compressed}. Additionally, recently, a dual-network architecture has been proposed for the joint estimation of sound speed and initial pressure from boundary measurements~\cite{hwang2025selfpat}. Other related works include spectral unmixing~\cite{cheng2024research} and unsupervised blind restoration in PAT~\cite{tang2023learning}.

In contrast to the aforementioned post-processing approaches, we embed DIP directly into a physics-based reconstruction framework, where the forward operator is included in the optimization objective, enforcing data fidelity in the \textit{measurement domain}. Additionally, instead of using a matrix representation, we make use of an analytic forward model, which enables high-resolution reconstructions.
While the related studies also incorporate the physical modeling, our focus is specifically on limited-view circular geometries as a result of detector subsampling and on the systematic evaluation of reconstruction quality under varying angular detector coverage and noise levels.

Approaches that include the model remain computationally demanding due to the repeated use of forward and adjoint models \cite{grohl2021deep,hauptmann2025fast,hauptmann2024model}, particularly for detection geometries other than linear and planar sensors. While fast inverse operators have existed for linear and planar acquisition geometries\cite{kostli2001temporal,cox2005fast,haltmeier2009reconstruction}, recently introduced fast algorithms for circular geometries~\cite{hauptmann2025fast} enable efficient implementation of iterative optimization frameworks such as DIP for the popular circular acquisition scheme. Consequently, we are able to evaluate the forward and adjoint at high spatial resolutions, which is essential when the operators are repeatedly called inside iterative optimization schemes such as DIP.  \\

In this work, we pursue a two-fold objective:
\begin{enumerate}[label=(\roman*)]
    \item to enable computationally efficient, high-resolution 2D PAT reconstruction by combining DIP with fast forward and adjoint operators for circular measurement geometries, and
    \item to systematically evaluate reconstruction performance under limited-view conditions induced by detector subsampling and measurement noise, using classical total variation (TV) reconstructions as a baseline.
\end{enumerate}

Our experiments include both simulated and experimental PAT data, including \emph{in-vivo} measurements. Our study suggests that the DIP framework provides a robust, unsupervised reconstruction strategy even under challenging acquisition conditions.

\section{Methods}

\subsection{Photoacoustic Tomography}

We remind that the ultrasound waves are detected by an array of ultrasonic transducers positioned around the region of interest. The measured acoustic data are used in the reconstruction of an image of the initial pressure distribution. In the following we will write the forward problem corresponding to \eqref{eqn:waveEq} and \eqref{eqn:measurements} by the usual operator equation
\begin{equation}
g^{\delta}=\mathcal{A}f + \delta,
\label{eq:InverseProblemEquation}
\end{equation}
where $\mathcal{A}$ is the forward operator describing the acoustic wave propagation \eqref{eqn:waveEq}, \textit{f} is the unknown image, $\delta$ is additive measurement noise, and $g^{\delta}$ is the noisy measurement.

\begin{figure}[h!tb]
\begin{center}
\includegraphics[width=0.5\linewidth]{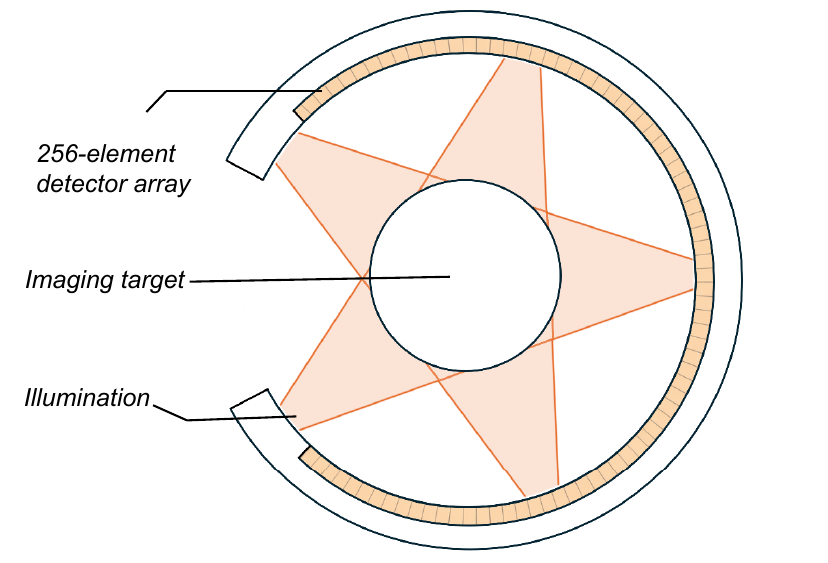}
\end{center}
\par
\vspace{-6mm}\caption{Schematic visualization of the MSOT inVision 256-TF by iThera Medical. The data are acquired by a ring array of 256 transducer elements that have an angular coverage of 270$^\circ$. The imaging target is positioned in the center of the ring array and is illuminated from five directions in 72$^\circ$ intervals, leading to a homogeneous radiant exposure in the imaging plane.
}
\label{fig:MSOTdevice}
\end{figure}

To reflect real-world limitations in sensor geometry, we investigate our approach under restricted angular coverage. That means the measurement domain $\Gamma_m$ only covers part of $\Omega$ and hence we will deal with a limited-view problem.  We conduct experiments using both simulated phantom data and experimental measurements, see Figure \ref{fig:MSOTdevice} for an illustration of the measurement system. We evaluate reconstruction quality across progressively reduced angular coverage. The experimental data examines three detector configurations with 256, and sub-sampled 170 and 112 detectors, resulting in angular coverages of $270^{\circ}$, $179.30^{\circ}$, and $118.125^{\circ}$, respectively. In the phantom studies, we focus on noise robustness by reconstructing images from $270^{\circ}$ simulated measurements while introducing normally distributed noise with different noise levels (0\%, 10\%, and 20 \%).

As a baseline, we compute classical total variation (TV) regularized reconstructions by minimizing the variational functional 
\begin{equation}
    \widehat{f} = \arg\min_{f \geq 0}\frac{1}{2}\|\mathcal{A}f-g^{\delta}\|_2^2 + \alpha\mathrm{TV}(f),
    \label{eq:classic tv}
\end{equation}
where $\alpha>0$ denotes the regularization weight. Reconstructions are computed with the iterative primal-dual hybrid gradient method \cite{chambolle2011first}. For experimental mouse data, the non-negativity constraint is removed and and additional quadratic penalty on the mean intensity added to encourage solutions with expected intensity level.

The computation time to obtain reconstructions is limited by the efficiency of the forward $\mathcal{A}$ and adjoint $\mathcal{A}^*$ operators. A general time-stepping approach is available in the popular k-Wave toolbox~\cite{treeby2010k}. The corresponding
forward and adjoint operators have been specifically studied to enable their use in iterative reconstruction methods\cite{haltmeier2017analysis,arridge2016adjoint}.  
Faster implementations can be obtained using a single step fast Fourier Transform (FFT) under the assumption of constant speed-of-sound. 
For planar and linear detector geometries, such efficient implementations have long been available, especially for the inverse\cite{kostli2001temporal,cox2005fast,haltmeier2009reconstruction}. 

Fast FFT-based reconstruction algorithms for circular and spherical acquisition geometries were also developed~\cite{kunyansky2012fast,kunyansky2007explicit,grohl2025digital}, providing explicit inversion formulas. However, fast forward and adjoint pairs for the popular 2D circular geometry were not introduced until recently~\cite{hauptmann2025fast}. 
The lack of fast enough forward and adjoint pairs for 2D circular geometry made the use of DIP, and other approaches that require repeated evaluation of forward and adjoint, impractical for higher resolutions. We employ these recently introduced fast FFT-based algorithms, which enable asymptotically fast evaluation of both $\mathcal{A}$ and $\mathcal{A}^*$ under the assumption of constant speed-of-sound. These models are essential to enable deep learning based methods, as we discuss next.

\subsection{Deep Image Prior}

In our work, Deep Image Prior (DIP) is employed as an unsupervised reconstruction strategy in which a convolutional neural network is optimized directly for the noisy and incomplete measurement, as described previously~\cite{baguer2020computed, barbano2022educated, dittmer2020regularization,ulyanov2018deep}. This approach relies on the inherent bias of the network architecture to guide the solution, which allows us to evaluate the effectiveness of such implicit regularization in limited-view PAT.

In the classic DIP approach, a neural network $\varphi_{\theta}$ takes a fixed input $z$ and produces an image. The forward operator $\mathcal{A}$ is then applied to this estimate, and the network parameters $\theta$ are optimized by minimizing the discrepancy between $\mathcal{A}(\varphi_{\theta}(z))$ and the noisy data $g^{\delta}$. Hence, there is a serious risk of overfitting to noise if training continues for too long. In other words, excessive emphasis on data fidelity may lead the network to reproduce noise artifacts present in the measurements. This motivates the use of an explicit regularization term in the loss function to encourage solutions that align with a priori assumptions about the solution. The balance between data fidelity and regularization is crucial: while the data fidelity term ensures consistency with the observed measurement, the regularization term imposes structural constraints that reflect prior knowledge of the underlying image~\cite{baguer2020computed, dittmer2020regularization}. 

The corresponding reconstruction problem using DIP for the inverse problem of Eq.~\eqref{eq:InverseProblemEquation} is then formulated as an optimization problem
\begin{equation}
\widehat{f}=\varphi_{\theta^*}(z) \quad \mathrm{ where }\quad \theta^*=\arg\min_{\theta}  \|\mathcal{A}\varphi_{\theta}(z)-g^{\delta}\|_2^2 + \lambda\mathrm{TV}(\varphi_{\theta}(z)).
\label{eq:DIP_total_minimization_with_TV}
\end{equation}
Here, $\lambda$ balances data fidelity and the TV regularization, which penalizes high-frequency components and mitigates overfitting. To further enhance reconstruction quality and accelerate convergence, we use an initial reconstruction $z=\mathcal{A}^{-1}g^{\delta}$ obtained with a fast FFT-based model for the inverse operator~\cite{hauptmann2025fast} rather than random noise, as in the classical DIP. The network takes the initial reconstruction as an input and outputs an updated image that, when passed through the forward operator, more closely matches the measured projections. This initialization provides a more informative starting point for optimization. With experimental mouse data, we include an additional quadratic penalty on the mean intensity of the reconstruction to encourage the solution to match the expected global intensity level. This acts as a soft constraint, stabilizing the optimization by removing uncertainty in the global intensity offset. Without this constraint, the optimization can produce reconstructions with a negative baseline drift.

The optimization process is carried out using the Adam optimizer with gradients computed via automatic differentiation to update the parameters $\theta$. As the forward operator $\mathcal{A}$ is present in the data fidelity term, the adjoint operator $\mathcal{A}^*$ is required to compute gradients. Repeated evaluations of the forward and adjoint operators form a computational bottleneck in learned iterative and optimization-based PAT reconstruction~\cite{hauptmann2025fast, hauptmann2024model}. This motivates the development of fast operator algorithms, as has been done for circular measurement geometry in our prior work~\cite{hauptmann2025fast}. Here, we test its use with the DIP reconstructions.

\subsubsection{Network architecture and training hyperparameters}

We employ a U-Net–based convolutional neural network \cite{ronneberger2015u} for the image reconstruction task. The neural network architecture follows the standard encoder-decoder structure with skip connections. The encoder contained four successive downsampling stages; each stage applied two $3 \times 3$ convolutional layers with batch normalization and ReLU activation. The number of feature channels was successively increased from 32 to 64, 128, and 256. Each downsampling stage applied $2\times2$ max pooling. 

The decoder symmetrically reverses the encoder module by three upsampling stages. Each of these stages employed a transposed convolution for upsampling, concatenation with the corresponding encoder feature maps via skip connections, and two consecutive $3 \times 3$ convolutional layers with batch normalization and ReLU activation. The final $3 \times 3$ convolution operation mapped the features to a single output channel. ReLU activation was applied to enforce non-negativity. For experimental mouse data, the final $3\times 3$ convolution was changed to a $1\times 1$ convolution without bias. Furthermore, the ReLU activation was changed to Leaky ReLU with 0.125 slope, as the measured mouse dataset exhibited approximately zero-centered amplitude distributions.

The neural network was trained using the Adam optimizer with a learning rate of $5 \cdot 10^{-4}$. The learning rate was adjusted using a cosine annealing schedule over the full number of iterations.

\subsection{Experimental Design}

For fair comparison, the regularization parameters $\alpha$ (TV) and $\lambda$ (DIP) are tuned via manual grid search for both the simulation study and experimental phantom data. Reconstructions are computed using different regularization parameters $\alpha$ and $\lambda$ for all samples. The best regularization parameter for each method and case is then chosen based on the overall image quality assessment (IQA) metric values with respect to the ground truth. For phantoms, this was done with respect to the whole image, and with experimental phantom data, this was restricted to the region of interest (ROI) consisting of the central circular area covering the object. As no ground truth was available for the experimental mouse data, the choice of regularization parameter $\lambda$ was guided by the visual assessment of artifact suppression and preservation of structural detail. The regularization parameter was selected empirically by evaluating a range of possible values and estimating their influence on the reconstruction quality.

Once tuned, the regularization parameters are kept fixed. As the ground truth is available for phantom simulations, the optimal iterate was selected based on PSNR. With early stopping DIP, the optimal iterate was chosen based on the globally highest PSNR value. For converged DIP, PSNR values for the first 40 iterations were discarded, after which the iteration producing the best PSNR was selected. For experimental phantom data, the reconstructions for both DIP and primal-dual are performed without using any ground truth information. In this unsupervised setting, neural network parameters for DIP are optimized only with respect to the given degraded measurements, as in \eqref{eq:DIP_total_minimization_with_TV}. 

With experimental phantom data, stopping criteria had to be selected independent of ground truth. For DIP, we selected a blunt maximum iteration cut-off of 400 iterations. This stopping point was determined by empirical analysis of PSNR (peak signal-to-noise ratio) curves across multiple samples. The selected iteration corresponds to a compromise that yielded generally high PSNR values while avoiding overfitting to noise observed in some samples at later stages of optimization. For some samples and angular setups, this cutoff happened before the PSNR optimum, while for others, the peak was achieved before reaching our stopping criterion. For the primal-dual algorithm, the final reconstruction was selected by tracking the relative difference between consecutive reconstructions, and accepted as converged if 
 $$\dfrac{\|f^{(k)}-f^{(k-1)}\|_2}{\|f^{(k)}\|_2} \leq 10^{-4}.$$

For experimental mouse data, the stopping criteria also had to be selected independently of ground truth. Based on qualitative evaluation, a stopping point at 600 iterations was selected as a compromise between limited-view artifact suppression and preservation of structural details.

\subsection{Data Sets}

\subsubsection{Simulations}

To avoid inverse crime, the forward problem is computed on a finer Cartesian grid (512×512) using an FFT-based wave propagation method, similar to the k-wave algorithm~\cite{treeby2010k}. The rest of the computational techniques use polar grid computations in the spectral domain on a coarser grid (256$\times$256).
To mimic a limited-view setup, measurements from detectors beyond the desired detectors are set to zero. Additionally, Gaussian noise is sampled and scaled to obtain desired noise levels.

\subsubsection{Experimental phantom data}

For the performance evaluation of our DIP method, we use a previously published digital twin dataset that combined experimental measurement data with a simulated reference of the initial pressure distribution~\cite{grohl2023moving,grohl2025digital}, allowing evaluation of the reconstruction quality. In brief, the data are experimentally captured with the MSOT InVision 256-TF with 270$^\circ$ angular coverage of the detection array, see Figure \ref{fig:MSOTdevice}. A full description of the device design can be found in the SIMPA framework~\cite{grohl2022simpa}. Tissue-mimicking test objects~\cite{grohl2023moving} were measured, whose optical properties are carefully characterized with a double-integrating sphere system and then used for a forward simulation of light transport. The method pipeline from measurement set-up to reconstructed image is illustrated in Figure \ref{fig:pa_exp_pipeline}. 

For the digital twin, simulations are done using SIMPA~\cite{grohl2022simpa}, which, in turn, uses the MCX Monte Carlo model~\cite{fang2009monte} for optical simulation and the k-Wave toolbox for acoustic simulations~\cite{treeby2010k}. The simulated and experimental images have a spatial resolution of 760$\times$760 pixels. For IQA evaluation and plotting, the images are cropped to 322$\times$322 pixels. In this study, we used the experimental measurements and compared the reconstruction results to the simulated initial pressure distributions as in \cite{grohl2025digital} to use full reference image-quality measures to compare different reconstruction schemes. We used sub-sets of the acoustic measurement data in the angular coverage dimension to emulate even stronger limited-view settings. The full 270$^{\circ}$ measurement data has a shape 256$\times$2030 (number of detectors$\times$number of time samples), while for the 179.3$^{\circ}$ and 118.12$^{\circ}$ cases, the detector dimension is subsampled to 170 and 112 detectors, respectively.

A binary region-of-interest (ROI) mask separating the phantom from the background was generated from the ground truth image and used to restrict all image-quality assessment (IQA) to the phantom region. Pixels outside the mask were zeroed out, and images were cropped for both ground truth and reconstruction before plotting and computing the IQA measures. 

\begin{figure}[t]
\centering
\resizebox{\textwidth}{!}{%
\begin{minipage}{0.15\textwidth}
    \centering
    \raisebox{0.25cm}{\includegraphics[height=2.0cm, width=2.0cm, clip]{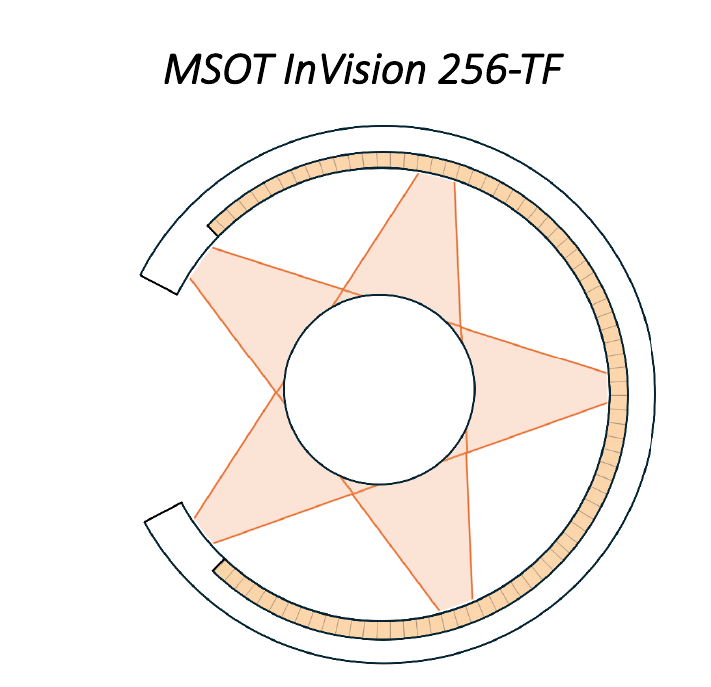}}\\
    \small Measurement set-up\\
    \phantom{}
\end{minipage}%
\raisebox{0.5cm}{$\xrightarrow{\substack{\text{data}\\ \text{acquisition}}}$}%
\begin{minipage}{0.14\textwidth}
    \centering
    \includegraphics[height=2cm,width=1.5cm,clip]{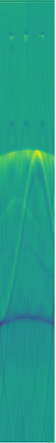}\\[4pt]
    \small Time-series data $p(x_s,t)$\\
    \phantom{ }
\end{minipage}%
\raisebox{0.5cm}{$\xrightarrow{\substack{\text{FFT}\\ \text{inverse}\\\mathcal{A}^{-1}}}$}%
\begin{minipage}{0.17\textwidth}
    \centering
    \includegraphics[height=2cm,width=2cm,clip]{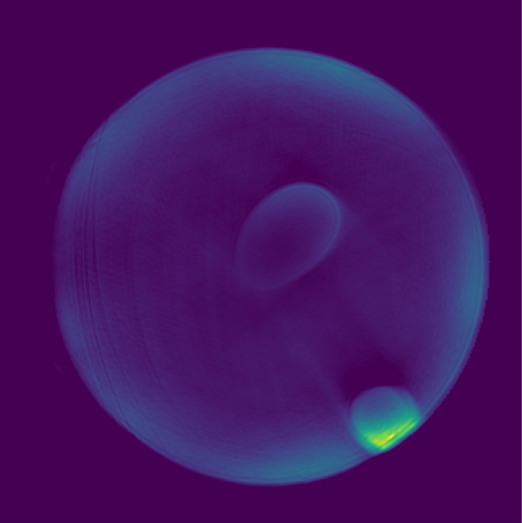}\\[4pt]
    \small Initial reconstruction\\ $z$
\end{minipage}%
\raisebox{0.5cm}{$\xrightarrow{\substack{\text{DIP}\\ \text{optimi-}\\ \text{zation}}}$}%
\begin{minipage}{0.18\textwidth}
    \centering
    \includegraphics[height=2cm,width=2cm,clip]{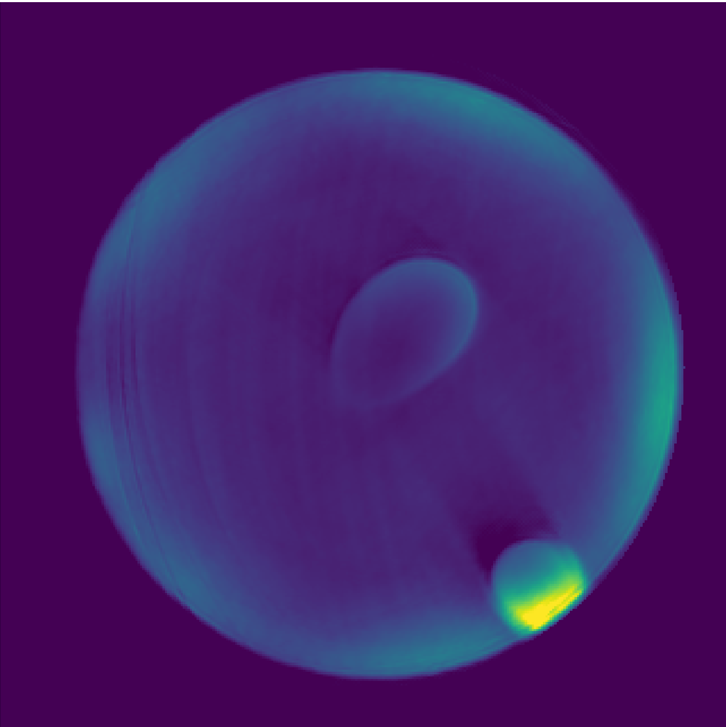}\\[4pt]
    \small Final reconstruction\\ $\widehat{f}=\varphi_{\theta^*}(z)$
\end{minipage}%
}
\caption{Method pipeline for experimental data. Time-series data is measured using a circular detection geometry and mapped to an initial pressure estimate using a fast inverse operator. This initial reconstruction is refined using the DIP framework, where an untrained U-Net is optimized by minimizing a data-fidelity term together with a total-variation regularization term. The final reconstruction $\widehat{f}$ is obtained as the output of the optimized network $\varphi_{\theta^*}$.}
\label{fig:pa_exp_pipeline}
\end{figure}

\subsubsection{In-vivo mouse data}

For a further qualitative performance evaluation of our DIP method, we use in vivo measurements of a mouse, obtained with the same measurement setup as the experimental phantom data. All animal procedures were conducted under project and personal licenses issued under the United Kingdom Animals (Scientific Procedures) Act, 1986, and compliance was approved locally by the CRUK Cambridge Institute Biological Resources Unit. The data were previously published by Gr\"ohl et al.~\cite{grohl2024distribution}. The mouse was lying with its spine towards the bottom of the detector array, producing transverse cross-sectional images of the mouse torso. 

The data was acquired at 40\,MHz sampling frequency using 256 detector elements with a 270$^{\circ}$ detector coverage, resulting in a data shape of 256$\times$2030 (number of detectors$\times$number of time samples). The maximum reconstructable area is 81$\times$81\,mm (in our case 760$\times$760 pixels), which was cropped to 32$\times$32\,mm (300$\times$300 pixels) for plotting.

\subsection{Evaluation Measures}

As discussed by Breger et al.\cite{breger2025study}, many commonly used full-reference image quality assessment (FR-IQA) methods were originally developed for \textit{natural images}, which have different structural and statistical properties than medical images, thus making their implementation often problematic. In the field of photoacoustic imaging, no standardized or tailored FR-IQA metrics currently exist, and therefore, conventional metrics such as the Structural Similarity Index (SSIM) and Peak Signal-to-Noise Ratio (PSNR) are often applied. As emphasized by the authors, both of these metrics can be misleading in medical settings when assessing the quality of a PA reconstruction. However, to enable comparison with prior work in the field, we display these metrics for our PAT reconstructions.  

Given the limitations of SSIM and PSNR, such as misjudgment between blur and sharpness, we further employ the Learned Perceptual Image Patch Similarity (LPIPS). Although LPIPS is not free of shortcomings, \cite{breger2025study}
 showed that it is able to handle small spatial misalignments more robustly compared to SSIM and PSNR, which is critical for PAT reconstructions. To further ensure balanced evaluation of image quality, we include HaarPSI - a Haar Wavelet-Based Perceptual Similarity Index – and the Pearson Correlation Coefficient (CC), which provides a complementary measure of structural agreement. 

 We compute SSIM and PSNR using implementations provided in the open-source image processing library \textit{scikit-image}. For both of these metrics, the dynamic range was set to the intensity range of the ground truth image within the ROI. With SSIM, we used default constants ($K_1$ = 0.01, $K_2$ = 0.03, no Gaussian weighting). With LPIPS \cite{zhang2018unreasonable}, we employed AlexNet, a CNN architecture, as the feature network. LPIPS was computed after standard inputs were scaled to [-1,1]. CC was computed on the flattened, nonzero ROI pixels. 

 The used IQA measures can be divided into two classes. LPIPS, measuring learned perceptual distance, is a distance metric with scale $[0, \infty)$, with a lower value indicating closer perceptual similarity. Other used metrics are similarity metrics, where a higher value indicates better image quality. SSIM measures structural similarity in a scale [0,1], PSNR measures MSE-based log-fidelity with a theoretical scale of $(-\infty, \infty)$. CC has a scale of $[-1, 1]$ and measures linear similarity between the reconstruction and the reference. Lastly, HaarPSI measures perceptual similarity based on Haar wavelet features in a scale $[0,1]$.

\section{Results}

\FloatBarrier
\subsection{Simulation study}


\newsavebox{\phantomgrid}
\sbox{\phantomgrid}{%
  \begin{tikzpicture}

  \node at (3.45,8.8) {\shortstack{\textbf{\footnotesize Initial}\\\textbf{\footnotesize reconstruction}}};
  \node at (6.05,8.8) {\shortstack{\textbf{\footnotesize DIP early }\\\textbf{\footnotesize stopping}}};
  \node at (8.65,8.8) {\shortstack{\textbf{\footnotesize DIP }\\\textbf{\footnotesize converged}}};
  \node at (11.25,8.8){\shortstack{\textbf{\footnotesize TV}\\\textbf{\footnotesize reconstruction}}};

  \node[rotate=90] at (1.4,7.15) {\shortstack{\textbf{\footnotesize 0\% relative}\\\textbf{\footnotesize L2-noise}}};
  \node[rotate=90] at (1.4,4.55) {\shortstack{\textbf{\footnotesize 10\% relative}\\\textbf{\footnotesize L2-noise}}};
  \node[rotate=90] at (1.4,1.95) {\shortstack{\textbf{\footnotesize 20\% relative}\\\textbf{\footnotesize L2-noise}}};

   \node[anchor=south west, inner sep=0] at (2.2,5.9)
      {\includegraphics[height=2.5cm]{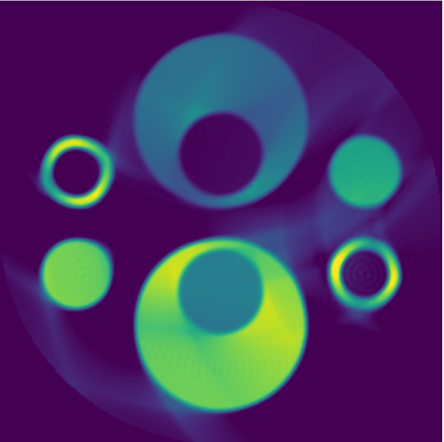}};
  \node[anchor=south west, inner sep=0] at (4.8,5.9)
      {\includegraphics[height=2.5cm]{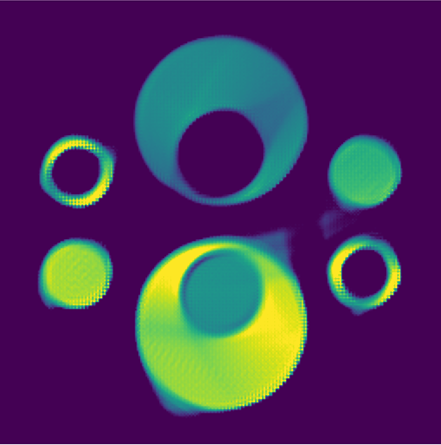}};
  \node[anchor=south west, inner sep=0] at (7.4,5.9)
      {\includegraphics[height=2.5cm]{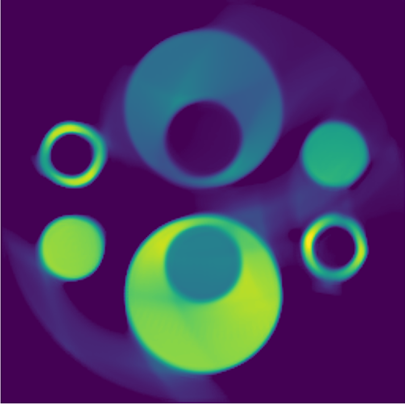}};
  \node[anchor=south west, inner sep=0] at (10.0,5.9)
      {\includegraphics[height=2.5cm]{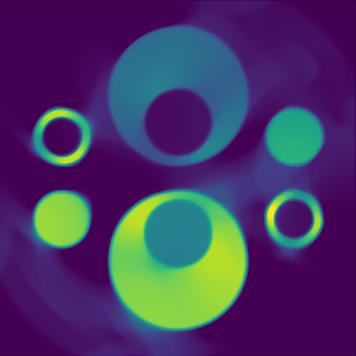}};

 \node[anchor=south west, inner sep=0] at (2.2,3.3)
      {\includegraphics[height=2.5cm]{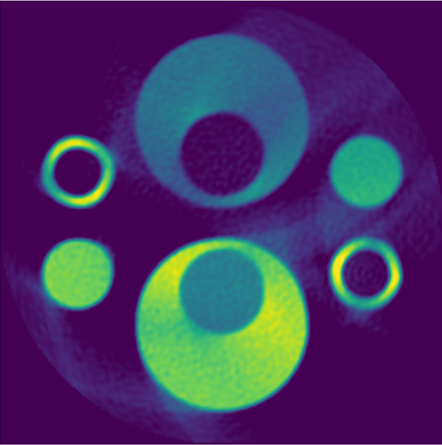}};
  \node[anchor=south west, inner sep=0] at (4.8,3.3)
      {\includegraphics[height=2.5cm]{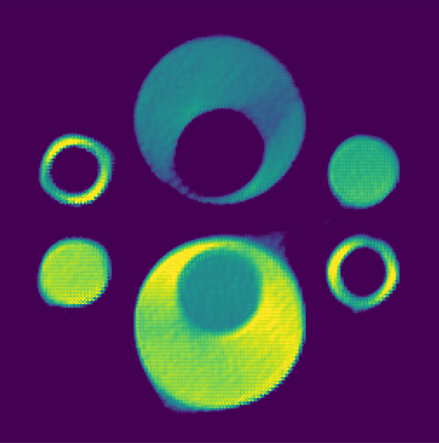}};
  \node[anchor=south west, inner sep=0] at (7.4,3.3)
      {\includegraphics[height=2.5cm]{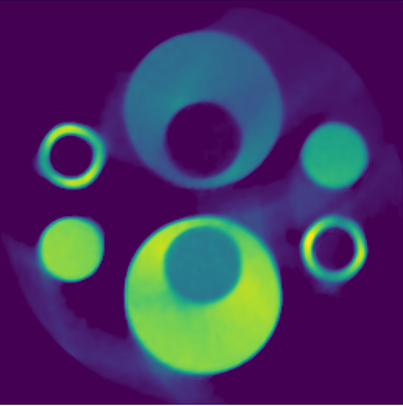}};
  \node[anchor=south west, inner sep=0] at (10.0,3.3)
      {\includegraphics[height=2.5cm]{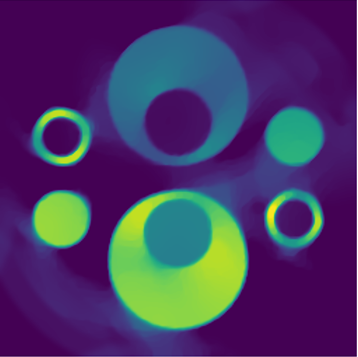}};

 \node[anchor=south west, inner sep=0] at (2.2,0.7)
      {\includegraphics[height=2.5cm]{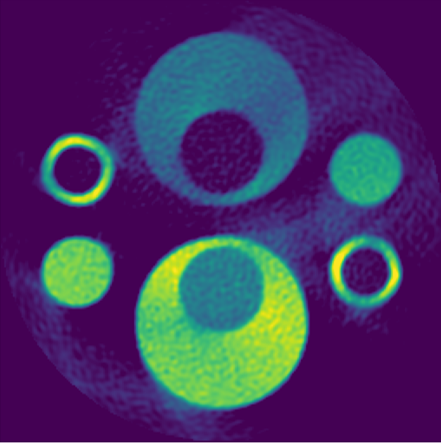}};
  \node[anchor=south west, inner sep=0] at (4.8,0.7)
      {\includegraphics[height=2.5cm]{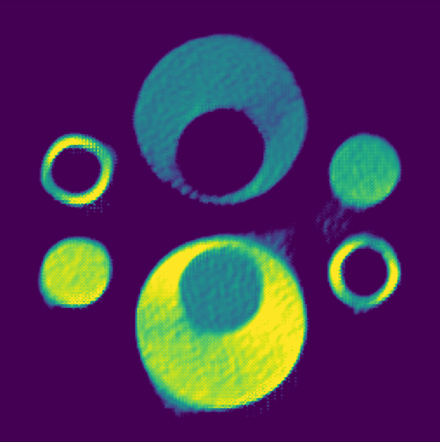}};
  \node[anchor=south west, inner sep=0] at (7.4,0.7)
      {\includegraphics[height=2.5cm]{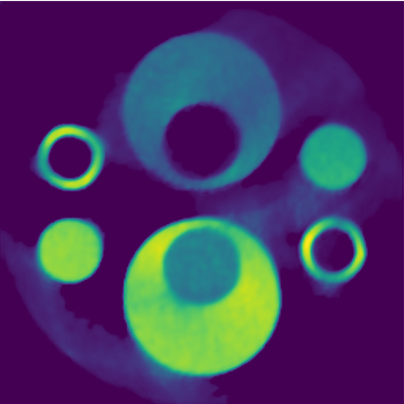}};
  \node[anchor=south west, inner sep=0] at (10.0,0.7)
      {\includegraphics[height=2.5cm]{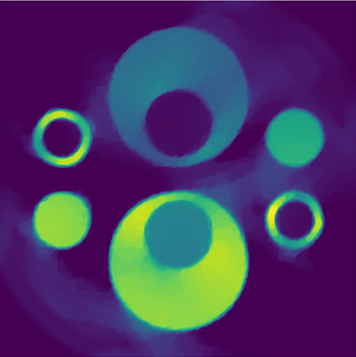}};

\end{tikzpicture}
}

\begin{figure}[t]
  \centering
  \begin{minipage}[t]{\wd\phantomgrid}
    \usebox{\phantomgrid}
  \end{minipage}%
  \hspace{0.0em}%
  \begin{minipage}[t]{0.03\textwidth}
  \raisebox{0.15cm}{
    \includegraphics[height=0.8\dimexpr\ht\phantomgrid+\dp\phantomgrid\relax,
      width=2\linewidth,
      clip,
      trim=0 2pt 0 2pt 
    ]{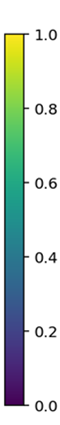}}
  \end{minipage}
  \caption{Comparison of reconstruction quality using DIP and TV regularization under 270° detector coverage at varying levels of added relative L2-noise (0\%, 10\%, and 20\%). Each row corresponds to a different noise level, while columns represent: (from left to right) initial reconstruction, DIP-based reconstruction with early stopping and running to convergence, and TV-regularized reconstruction.}
  \label{fig:phantom_noise_comparison}
\end{figure}

\begin{table}[h!]
    \footnotesize
    \centering
    \setlength{\tabcolsep}{10pt}   
    \renewcommand{\arraystretch}{1.5} 
    \setlength{\extrarowheight}{4pt}  

    \begin{tabular}{p{1.5cm} p{2.2cm}|cccc}
        \makecell{\textbf{L2-noise}}
        & \makecell{\textbf{IQA}}
        & \makecell{\textbf{Initial}\\ \textbf{(FFT)} } 
        & \makecell{\textbf{DIP early}\\ \textbf{stopping} }
        & \makecell{\textbf{DIP}\\ \textbf{converged}}
        & \makecell{\textbf{TV} } \\
        \hline

        \makecell[l]{\textbf{0\%} } 
        & \makecell[l]{\rule{0pt}{3.2ex}$\uparrow$ PSNR (dB)  \\ $\uparrow$ SSIM\\ $\downarrow$ LPIPS \\ $\uparrow$ CC \\ $\uparrow$ HaarPSI } 
        & \makecell[l]{\rule{0pt}{3.2ex}18.475  \\ 0.7110 \\  \textbf{0.079} \\  \textbf{0.970}\\ 0.533}
        & \makecell[l]{\rule{0pt}{3.2ex}\textbf{19.316} \\  \textbf{0.839   }\\  0.135  \\  0.961\\ 0.427}
        & \makecell[l]{\rule{0pt}{3.2ex}18.598 \\ 0.714 \\0.097 \\0.969 \\0.531}
        & \makecell[l]{\rule{0pt}{3.2ex}18.742\\  0.626 \\  0.084\\ 0.969\\\textbf{0.540}} \\
        \hline

        \makecell[l]{\textbf{10\%} } 
        & \makecell[l]{\rule{0pt}{3.2ex}$\uparrow$ PSNR (dB)   \\ $\uparrow$ SSIM \\ $\downarrow$ LPIPS \\ $\uparrow$ CC \\ $\uparrow$ HaarPSI} 
        & \makecell[l]{\rule{0pt}{3.2ex}18.432  \\ 0.677 \\ 0.145 \\ \textbf{0.969}\\ 0.506}
        & \makecell[l]{\rule{0pt}{3.2ex}\textbf{19.155}  \\ \textbf{0.813} \\ 0.159 \\ 0.965\\0.404}
        & \makecell[l]{\rule{0pt}{3.2ex}18.490 \\ 0.726 \\ 0.103 \\0.969 \\0.523}
        & \makecell[l]{\rule{0pt}{3.2ex}18.722  \\ 0.625 \\ \textbf{0.089}\\0.968\\\textbf{0.537}} \\
        \hline

        \makecell[l]{\textbf{20\%} } 
        & \makecell[l]{\rule{0pt}{3.2ex}$\uparrow$ PSNR (dB)   \\ $\uparrow$ SSIM \\ $\downarrow$ LPIPS \\ $\uparrow$ CC \\ $\uparrow$ HaarPSI} 
        & \makecell[l]{\rule{0pt}{3.2ex}18.323 \\ 0.616 \\ 0.228 \\ 0.967\\0.452}
        & \makecell[l]{\rule{0pt}{3.2ex}\textbf{19.368}  \\  \textbf{0.783} \\ 0.175  \\ 0.962\\0.407 }
        & \makecell[l]{\rule{0pt}{3.2ex}18.510 \\ 0.714 \\ 0.121 \\\textbf{0.968} \\ 0.507}
        & \makecell[l]{\rule{0pt}{3.2ex}18.678 \\ 0.627 \\ \textbf{0.113} \\ 0.968\\\textbf{0.528}} \\
        \hline

    \end{tabular}
    \caption{IQA measures (PSNR, SSIM, LPIPS, CC, HaarPSI) w.r.t.\ ground truth using DIP and TV regularization under 270° detector coverage with increasing relative L2-noise levels (0\%, 10\%, 20\%). All IQA measures are computed over the whole image. The arrows indicate the direction of improvement for each IQA metric, i.e., does a smaller or bigger value imply better image quality.}
    \label{table:table_of_errors}
\end{table}

\begin{figure}
\centering
\begin{minipage}{5.5cm}
  \includegraphics[width=4.5cm]{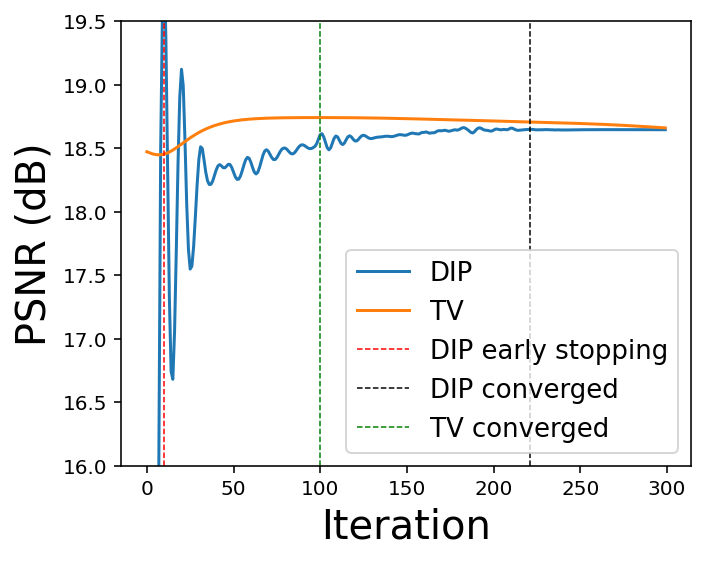}
  \caption*{(a) Noise-free case}
\end{minipage}
\qquad
\begin{minipage}{5.0cm}
\includegraphics[width=4.5cm]{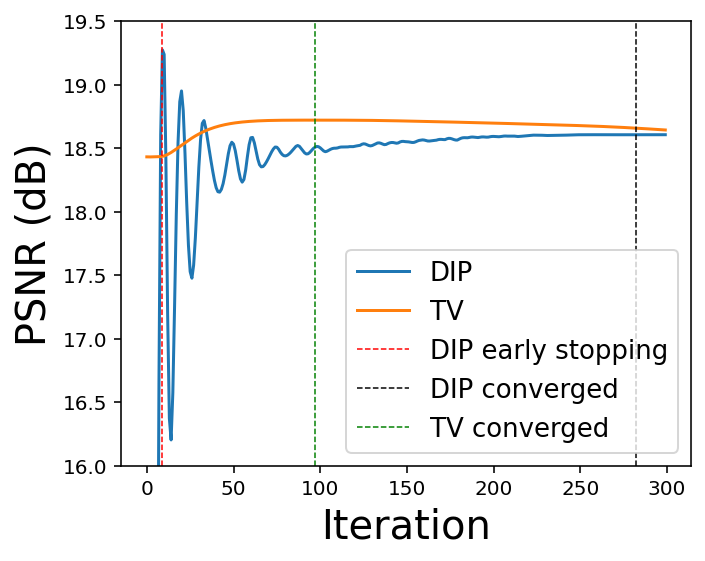}
\caption*{(b) 10\% noise}
\end{minipage}
\qquad
\begin{minipage}{5.0cm}
    \includegraphics[width=4.5cm]{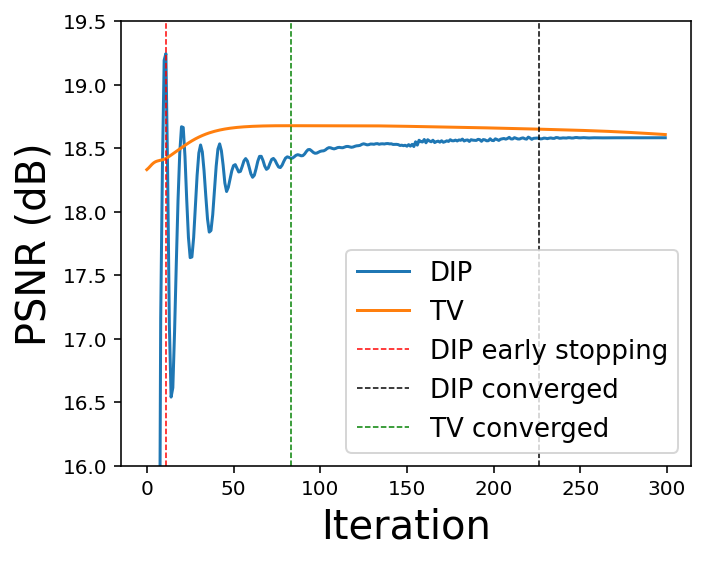}
    \caption*{(c) 20\% noise}
\end{minipage}
\caption{PSNR during reconstructions as a function of iteration of DIP and TV method under varying noise: (a) noise-free case, (b) 10\% added noise, and (c) 20\% added noise. Vertical lines indicate the early-stopping point for DIP and the convergence points for both of the methods. }
\label{fig:PSNR_sim}
\end{figure}

\begin{figure}
    \centering
    \includegraphics[width=0.5\linewidth]{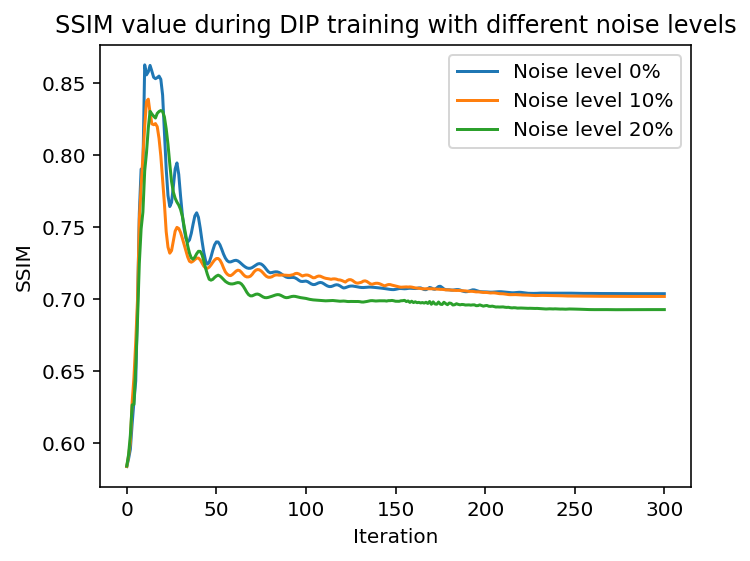}
    \caption{SSIM during DIP training under different noise levels (0\%, 10\% and 20\%). The curves show rapid improvement in the early iterations, followed by a gradual stabilization. Higher noise levels result in earlier degradation of performance and lower peak SSIM value.}
    \label{fig:SSIM_all_noise_levels}
\end{figure}

Results for the phantom are shown in Figure~\ref{fig:phantom_noise_comparison} and Table~\ref{table:table_of_errors} for varying levels of added relative L2 noise (0\%, 10\%, and 20\%). 
For TV reconstructions, we used $\alpha=0.05$ for all noise levels, 100 iterations in the noise-free case, 97 iterations with the 10\% case, and 83 iterations in the noisiest 20\% case. With the DIP framework, total variation regularization was applied with $\lambda=0.1$ for all noise cases. For DIP, we present two cases, one with early stopping and one at a converged state, to compare different image characteristics. Here, the early stopping corresponds to the iteration that generates the highest PSNR value. 
With the noise-free case, DIP with early stopping used 14 iterations, while converged DIP required 221 iterations. For 10\% and 20\% noise cases, these numbers were 11 and 282, 12 and 226, respectively. The DIP framework was initialized with FFT inverse.

Figure~\ref{fig:PSNR_sim} presents the PSNR progression for DIP and TV under noise-free, 10\% noise, and 20\% noise. For all noise levels, DIP exhibits a rapid increase in PSNR during the early iterations, followed by oscillatory behavior before stabilization. In contrast, TV demonstrates a smoother convergence pattern. Across all noise levels, DIP with early topping achieves the highest PSNR values, as also presented in Table~\ref{table:table_of_errors}. Fully converged DIP and TV reach similar but slightly lower PSNR values compared to DIP with early stopping. Figure~\ref{fig:SSIM_all_noise_levels} tells a similar story on DIP's behavior by illustrating the SSIM evolution during DIP training for different noise levels. For all three noise levels, SSIM increases rapidly in the early iterations, reaching a maximum before gradually decreasing as the iterations go on. Higher noise levels seem to reduce the peak SSIM value and accelerate the following decline, which indicates increased sensitivity to overfitting under noisier contexts.

As can be seen from Figure~\ref{fig:phantom_noise_comparison}, across all noise levels, the initial FFT reconstruction exhibits noticeable limited-view artifacts in the background, while the phantom itself remains largely intact with preserved shape and boundaries; however, increasing noise introduces visible graininess to the structures. DIP with early stopping substantially suppresses these background artifacts, yielding the cleanest reconstructions and maintaining well-defined object geometry even at higher noise levels. Conversely, TV and DIP run to convergence and produce qualitatively comparable images. Both methods sharpen the phantom boundaries but noticeably oversmooth high-frequency details. Unlike early-stopping DIP, however, neither TV nor converged DIP succeeds in removing the background artifacts, which remain visible across all noise levels.

The quantitative metrics in Table~\ref{table:table_of_errors} support the qualitative observations. Across all noise levels, the initial FFT reconstruction maintains reasonable structural accuracy, which is consistent with the well-preserved phantom shape, yet its PSNR and SSIM remain limited due to limited-view artifacts and increased graininess. DIP with early stopping consistently achieves the highest PSNR and SSIM among all methods. In contrast, TV and DIP at convergence yield similar quantitative values. They both maintain sharp phantom boundaries but exhibit reduced SSIM and PSNR compared to DIP with early stopping, which is consistent with their tendency to oversmooth high-frequency details. TV achieves the best LPIPS and HaarPSI values across all noise levels, although differences to converged DIP are insignificant. This indicates that TV and converged DIP provide better perceived sharpness and local structural fidelity. DIP with early stopping, despite its excellent global distortion measures, performs worst in LPIPS and HaarPSI. Overall, quantitative evaluation highlights a tradeoff: DIP with early stopping optimizes global similarity measures (PSNR and SSIM) while TV and converged DIP display improved perceptual sharpness.

\FloatBarrier
\subsection{Experimental phantom data and digital twin}

\subsubsection{256/256 detectors in use ($270^{\circ}$ angular coverage)}

\begin{figure}[h!]
    \centering
    
    \begin{minipage}[c]{0.08\textwidth}
        \centering
        \rotatebox{90}{
        \raisebox{0.0cm}{\scriptsize \textbf{Sample 1}}}
    \end{minipage}
    \begin{minipage}[c]{0.90\textwidth}
        \begin{minipage}[c]{0.2\textwidth}
            \centering
            \scriptsize \textbf{Ground truth \\ \phantom{}}\\
            \vspace{0.2em}
            \rotatebox{0}{
             \includegraphics[height=2.5cm, width=2.5 cm, clip]{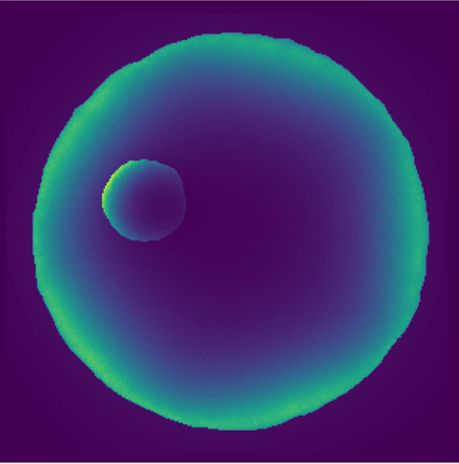}}\\
        \end{minipage}
        \hspace{0.002\textwidth}
        \begin{minipage}[c]{0.2\textwidth}
            \centering
            \scriptsize \textbf{Initial reconstruction}\\
            \vspace{0.2em}
            \rotatebox{0}{
            \includegraphics[height=2.5cm, width=2.5cm, clip]{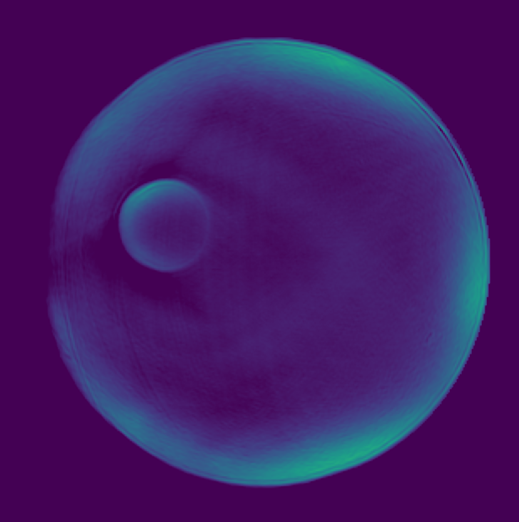}}\\

        \end{minipage}
        \hspace{0.002\textwidth}
        \begin{minipage}[c]{0.2\textwidth}
            \centering
            \scriptsize \textbf{DIP \\ \phantom{}}\\
            \vspace{0.2em}
            \rotatebox{0}{
            \includegraphics[height=2.5cm, width=2.5 cm, clip]{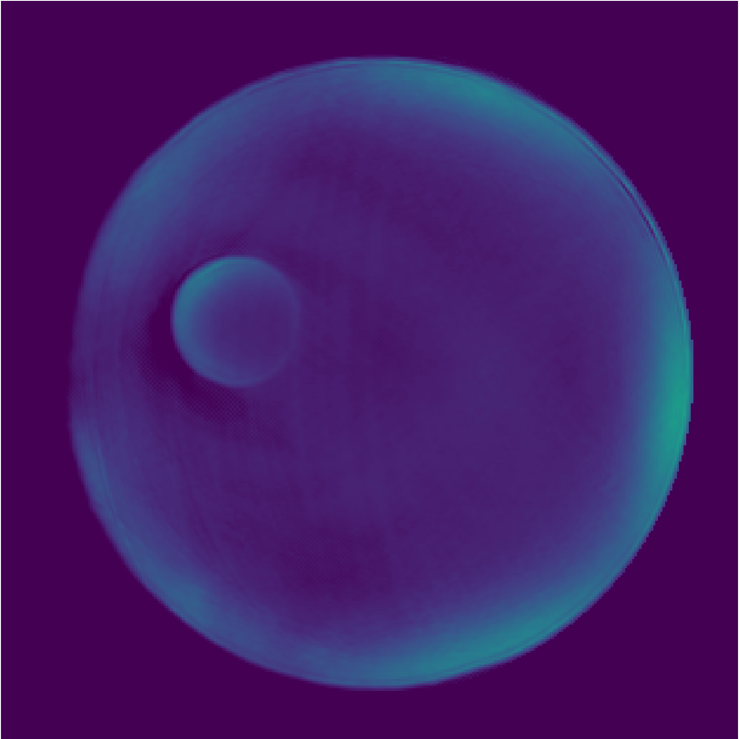}}\\

        \end{minipage}
        \hspace{0.002\textwidth}
        \begin{minipage}[c]{0.2\textwidth}
            \centering
            \scriptsize \textbf{TV \\ \phantom{}}\\
            \vspace{0.2em}
            \rotatebox{0}{
            \includegraphics[height=2.5cm, width=2.5cm, clip]{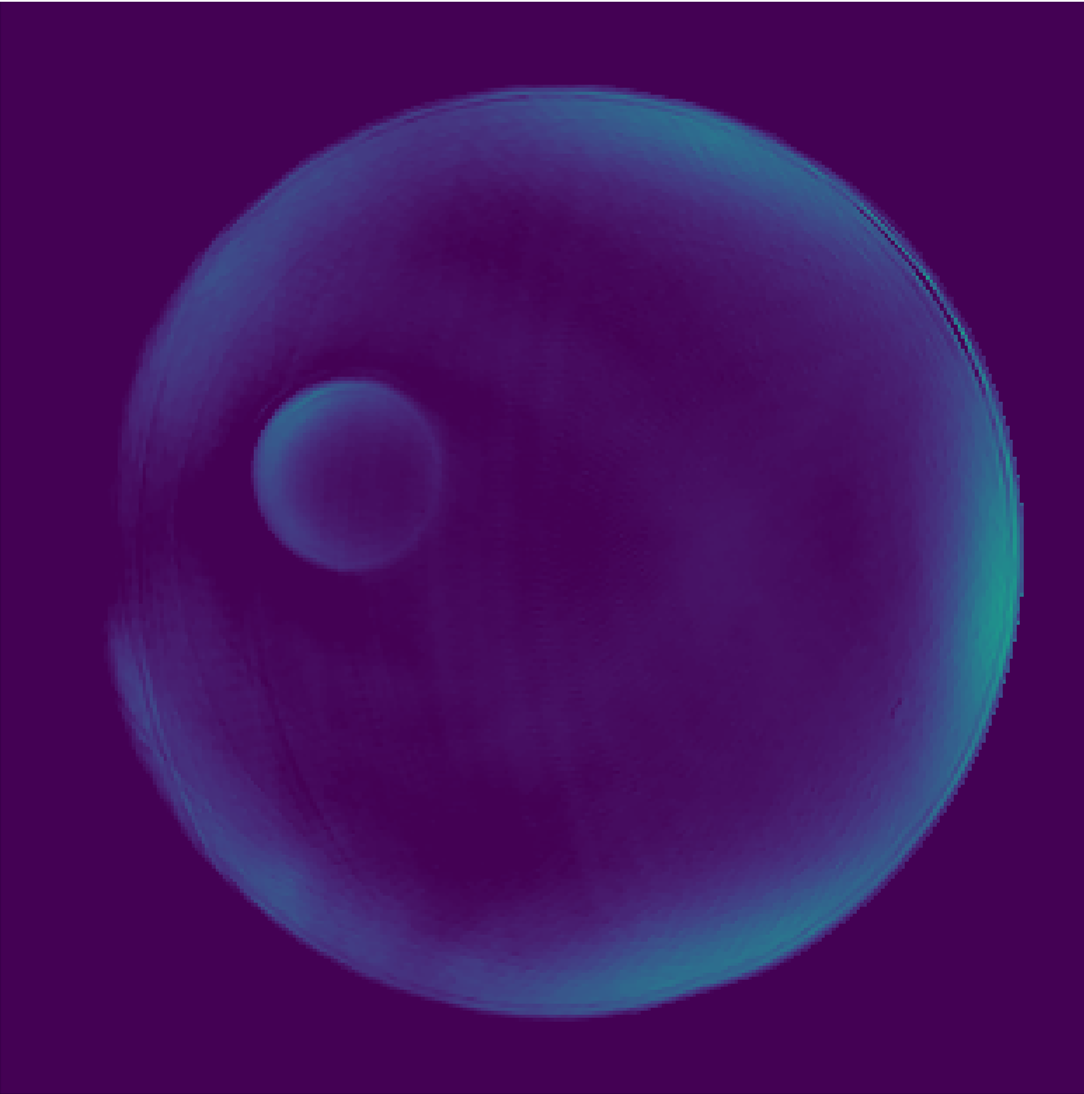}}\\

        \end{minipage}
        \begin{minipage}[c]{0.09\textwidth}
            \centering
            
            \vspace*{0.5cm}\includegraphics[height=2.5cm]{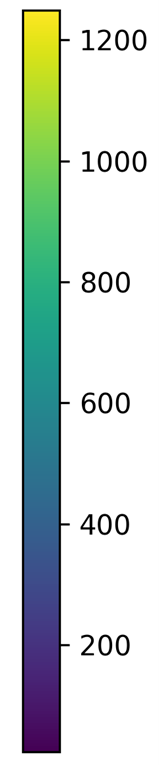}

        \end{minipage}
    \end{minipage}

    \vspace{1em}

    \begin{minipage}[c]{0.08\textwidth}
        \centering
        \rotatebox{90}{
        \raisebox{0.0cm}{\scriptsize \textbf{Sample 2}}}
    \end{minipage}
    \begin{minipage}[c]{0.90\textwidth}
        \begin{minipage}[c]{0.2\textwidth}
            \centering
            \rotatebox{0}{
            \includegraphics[height=2.5cm, width=2.5cm, clip]{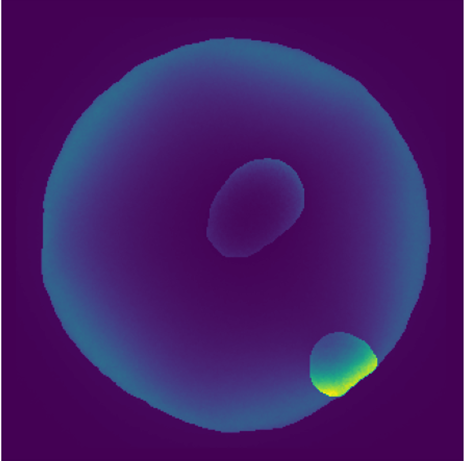}}\\
            \scriptsize
        \end{minipage}
        \hspace{0.002\textwidth}
        \begin{minipage}[c]{0.2\textwidth}
            \centering
            \rotatebox{0}{
            \includegraphics[height=2.5cm, width=2.5cm, clip]{images/initial_270_P.5.6.2_750.png}}\\

        \end{minipage}
        \hspace{0.002\textwidth}
        \begin{minipage}[c]{0.2\textwidth}
            \centering
            \rotatebox{0}{
            \includegraphics[height=2.5cm, width=2.5cm, clip]{images/DIP_270_P.5.6.2_750.png}}\\

        \end{minipage}
        \hspace{0.002\textwidth}
        \begin{minipage}[c]{0.2\textwidth}
            \centering
            \rotatebox{0}{
            \includegraphics[height=2.5cm, width=2.5cm, clip]{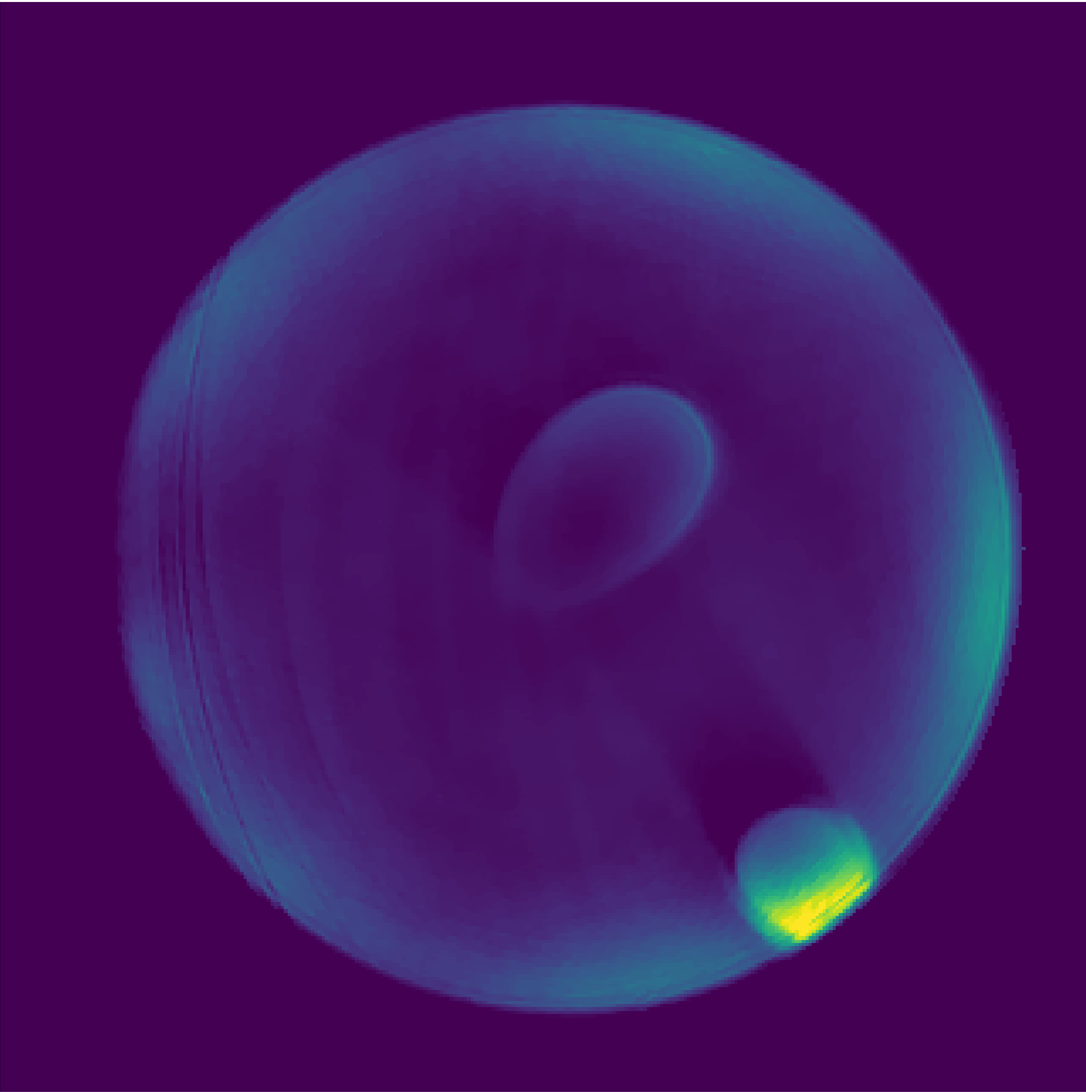}}\\
        \end{minipage}
        \begin{minipage}[c]{0.09\textwidth}
            \centering
            \hspace{0.125cm}\includegraphics[height=2.5cm]{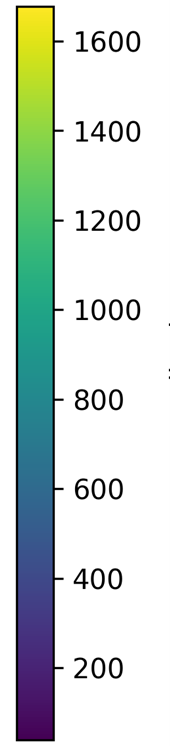}
        \end{minipage}
    \end{minipage}

    \caption{Comparison of reconstruction quality with all 256 detectors in use (270$^{\circ}$ angular coverage). }
    \label{fig:270_comparison}
\end{figure}

\begin{table}[h!]
    \footnotesize
    \centering
    \setlength{\tabcolsep}{10pt}   
    \renewcommand{\arraystretch}{1.5} 
    \setlength{\extrarowheight}{4pt}  

    \begin{tabular}{p{1.5cm} p{2.2cm}|ccc}
        \makecell{}
        & \makecell{\textbf{IQA}}
        & \makecell{\textbf{Initial}\\ \textbf{(FFT)} } 
        & \makecell{\textbf{DIP} }
        & \makecell{\textbf{TV} } \\
        \hline

        \makecell[l]{\textbf{Sample 1} } 
        & \makecell[l]{\rule{0pt}{3.2ex}$\uparrow$ PSNR (dB)  \\ $\uparrow$ SSIM\\ $\downarrow$ LPIPS \\ $\uparrow$ CC \\ $\uparrow$ HaarPSI } 
        & \makecell[l]{\rule{0pt}{3.2ex}14.016 \\ 0.632\\  0.309 \\  \textbf{0.702}\\ 0.223}
        & \makecell[l]{\rule{0pt}{3.2ex}\textbf{15.333}\\  \textbf{0.711}\\ \textbf{ 0.257} \\  0.680\\ \textbf{0.272}}
        & \makecell[l]{\rule{0pt}{3.2ex}14.422 \\ 0.649 \\0.273 \\0.693 \\0.250}
        \\
        \hline

        \makecell[l]{\textbf{Sample 2} } 
        & \makecell[l]{\rule{0pt}{3.2ex}$\uparrow$ PSNR (dB)   \\ $\uparrow$ SSIM \\ $\downarrow$ LPIPS \\ $\uparrow$ CC \\ $\uparrow$ HaarPSI} 
        & \makecell[l]{\rule{0pt}{3.2ex}21.101  \\ 0.779 \\ 0.253 \\0.763 \\ 0.318}
        & \makecell[l]{\rule{0pt}{3.2ex}\textbf{23.062}  \\\textbf{ 0.826} \\ \textbf{0.223} \\ 0.756\\\textbf{0.345}}
        & \makecell[l]{\rule{0pt}{3.2ex}21.888 \\ 0.796 \\ 0.223 \\\textbf{0.771} \\0.332}

    \end{tabular}
    \caption{IQA measures (PSNR, SSIM, LPIPS, CC, HaarPSI) w.r.t.\ ground truth using DIP and TV regularization under 270° detector coverage. All IQA measures are computed over ROI. The arrows indicate the direction of improvement for each IQA metric, i.e., does a smaller or bigger value imply better image quality.}
    \label{table:270}
\end{table}

The results for the full detector geometry (270$^{\circ}$) are shown in Figure \ref{fig:270_comparison} and Table \ref{table:270}. DIP uses $\lambda=0.075$ and $\lambda=0.75$ for Sample 1 and Sample 2, respectively.  Presented reconstructions were obtained after 400 iterations. Classic TV uses $\alpha=0.1$ and 266 iterations for Sample 1, and $\alpha=0.25$ and 220 iterations for Sample 2. All IQA metrics are computed only within the ROI.

Under full detector availability and 270$^{\circ}$ coverage, the initial reconstructions exhibit only mild limited-view artifacts, and the overall object geometry is well preserved in both samples. DIP and TV both further improve the image quality, and their reconstructions appear visually similar. However, with Sample 1, DIP achieves noticeably sharper edges in the inner circular boundary. In Sample 2, DIP again preserves the boundary of the central inclusion better and yields a brighter reconstruction overall. The qualitative differences are reflected in the IQA metrics, in which DIP consistently outperforms both the initial reconstruction and TV reconstruction in all other metrics except for the Pearson correlation coefficient (CC).

\subsubsection{170/256 detectors in use ($179.3^{\circ}$ angular coverage)}

\begin{figure}[h!]
    \centering

    \begin{minipage}[c]{0.08\textwidth}
        \centering
        \rotatebox{90}{
        \raisebox{0.0cm}{\scriptsize \textbf{Sample 1}}}
    \end{minipage}
    \begin{minipage}[c]{0.90\textwidth}
        \begin{minipage}[c]{0.2\textwidth}
            \centering
            \scriptsize \textbf{Ground truth \\ \phantom{}}\\
            \vspace{0.2em}
            \rotatebox{0}{
            \includegraphics[height=2.5cm, width=2.5cm, clip]{images/ground_truth_P.5.32_750.png}}\\
        \end{minipage}
        \hspace{0.002\textwidth}
        \begin{minipage}[c]{0.2\textwidth}
            \centering
            \scriptsize \textbf{Initial reconstruction}\\
            \vspace{0.2em}
            \rotatebox{0}{
            \includegraphics[height=2.5cm, width=2.5cm, clip]{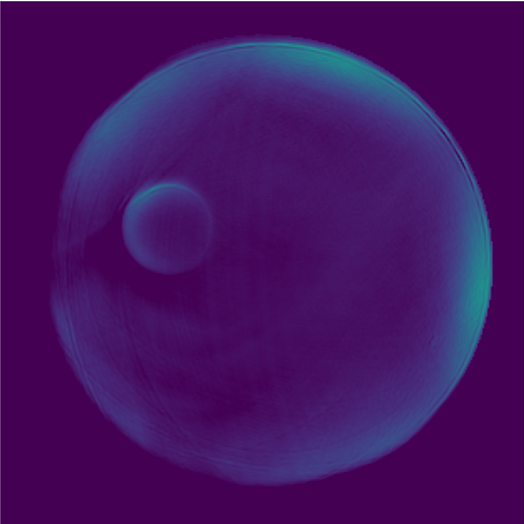}}\\

        \end{minipage}
        \hspace{0.002\textwidth}
        \begin{minipage}[c]{0.2\textwidth}
            \centering
            \scriptsize \textbf{DIP \\ \phantom{}}\\
            \vspace{0.2em}
            \rotatebox{0}{
            \includegraphics[height=2.5cm, width=2.5cm, clip]{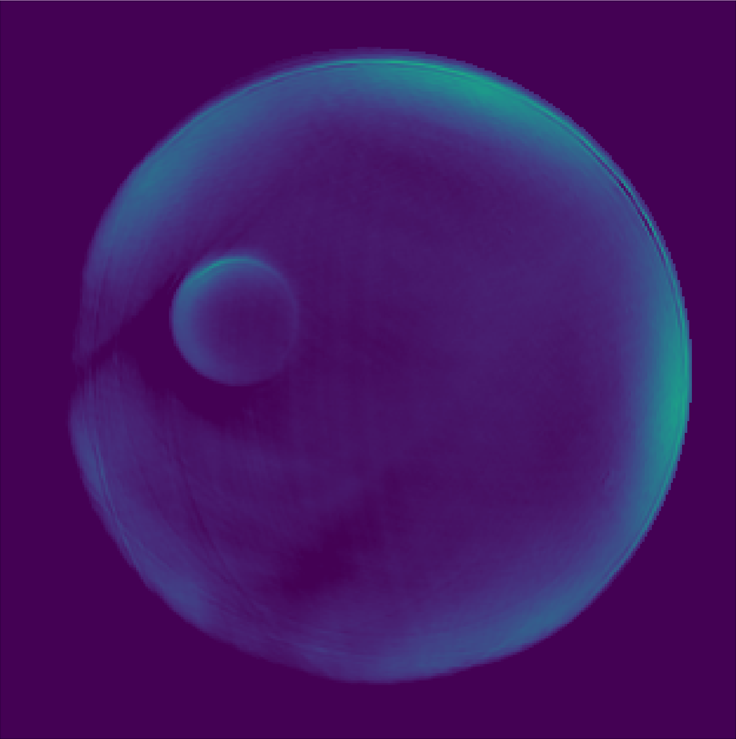}}\\

        \end{minipage}
        \hspace{0.002\textwidth}
        \begin{minipage}[c]{0.2\textwidth}
            \centering
            \scriptsize \textbf{TV \\ \phantom{}}\\
            \vspace{0.2em}
            \rotatebox{0}{
            \includegraphics[height=2.5cm, width=2.5cm, clip]{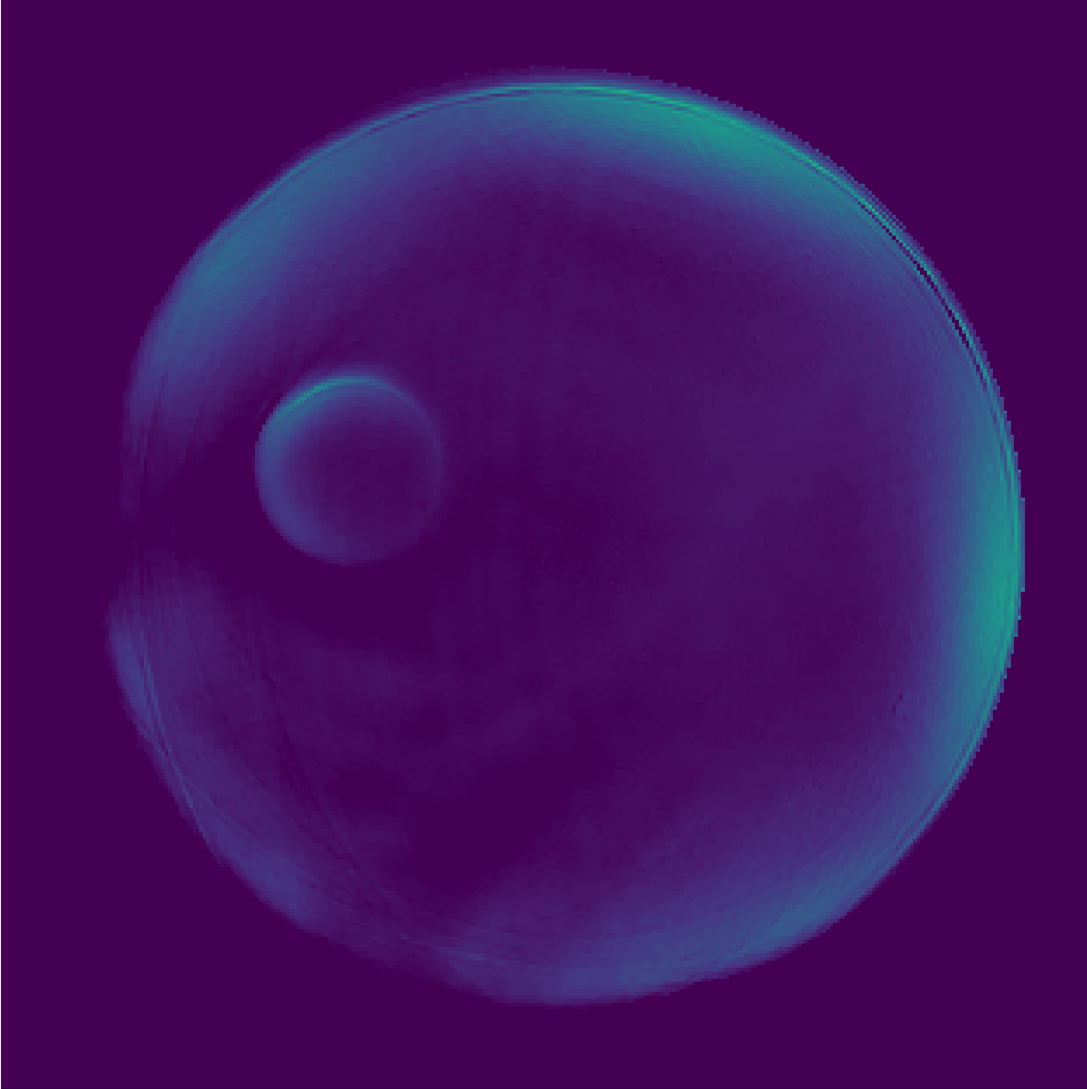}}\\
        \end{minipage}
        \begin{minipage}[c]{0.09\textwidth}
            \centering
            \vspace*{0.5cm}\includegraphics[height=2.5cm]{images/scale_P.5.32_750.png}
            
        \end{minipage}
    \end{minipage}

    \vspace{0.5em}

    \begin{minipage}[c]{0.08\textwidth}
        \centering
        \rotatebox{90}{\raisebox{0.0cm}{\scriptsize \textbf{Sample 2}}}
    \end{minipage}
    \begin{minipage}[c]{0.90\textwidth}
        \begin{minipage}[c]{0.2\textwidth}
            \centering
            \rotatebox{0}{
            \includegraphics[height=2.5cm, width=2.5cm, clip]
            {images/ground_truth_P.5.6.2_750.png}}\\
        \end{minipage}
        \hspace{0.002\textwidth}
        \begin{minipage}[c]{0.2\textwidth}
            \centering
            \rotatebox{0}{
            \includegraphics[height=2.5cm, width=2.5cm, clip]
            {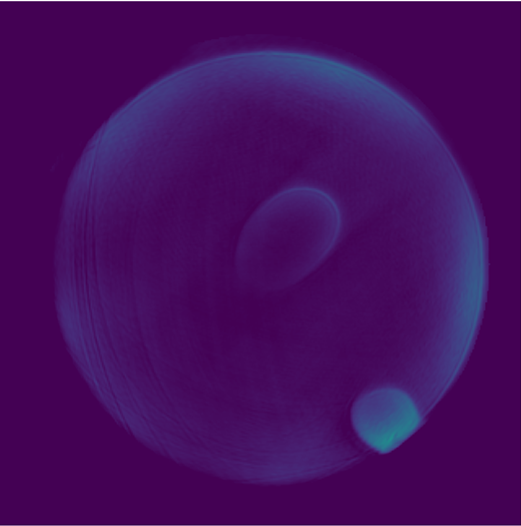}}\\

        \end{minipage}
        \hspace{0.002\textwidth}
        \begin{minipage}[c]{0.2\textwidth}
            \centering
            \rotatebox{0}{
            \includegraphics[height=2.5cm, width=2.5cm, clip]
            {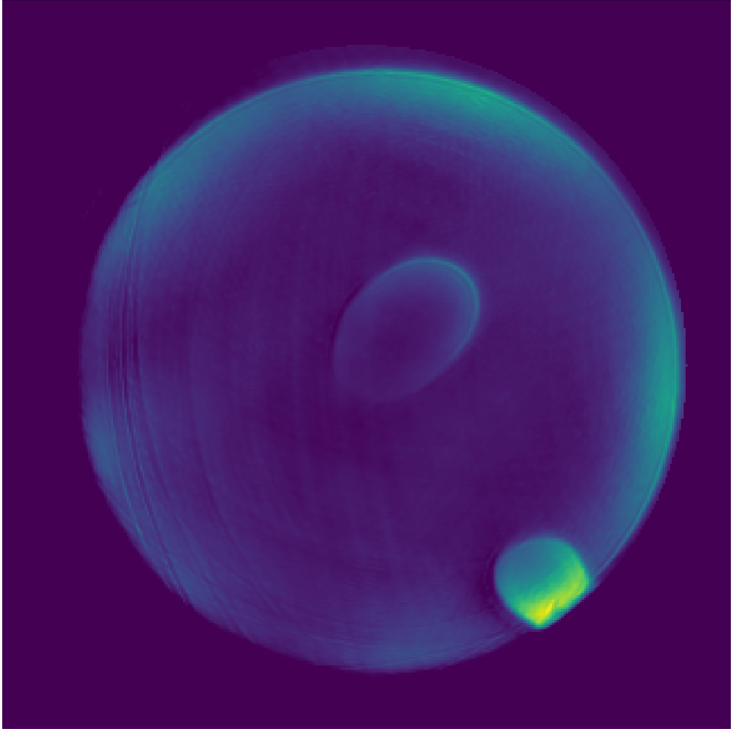}}\\

        \end{minipage}
        \hspace{0.002\textwidth}
        \begin{minipage}[c]{0.2\textwidth}
            \centering
            \rotatebox{0}{
            \includegraphics[height=2.5cm, width=2.5cm, clip]{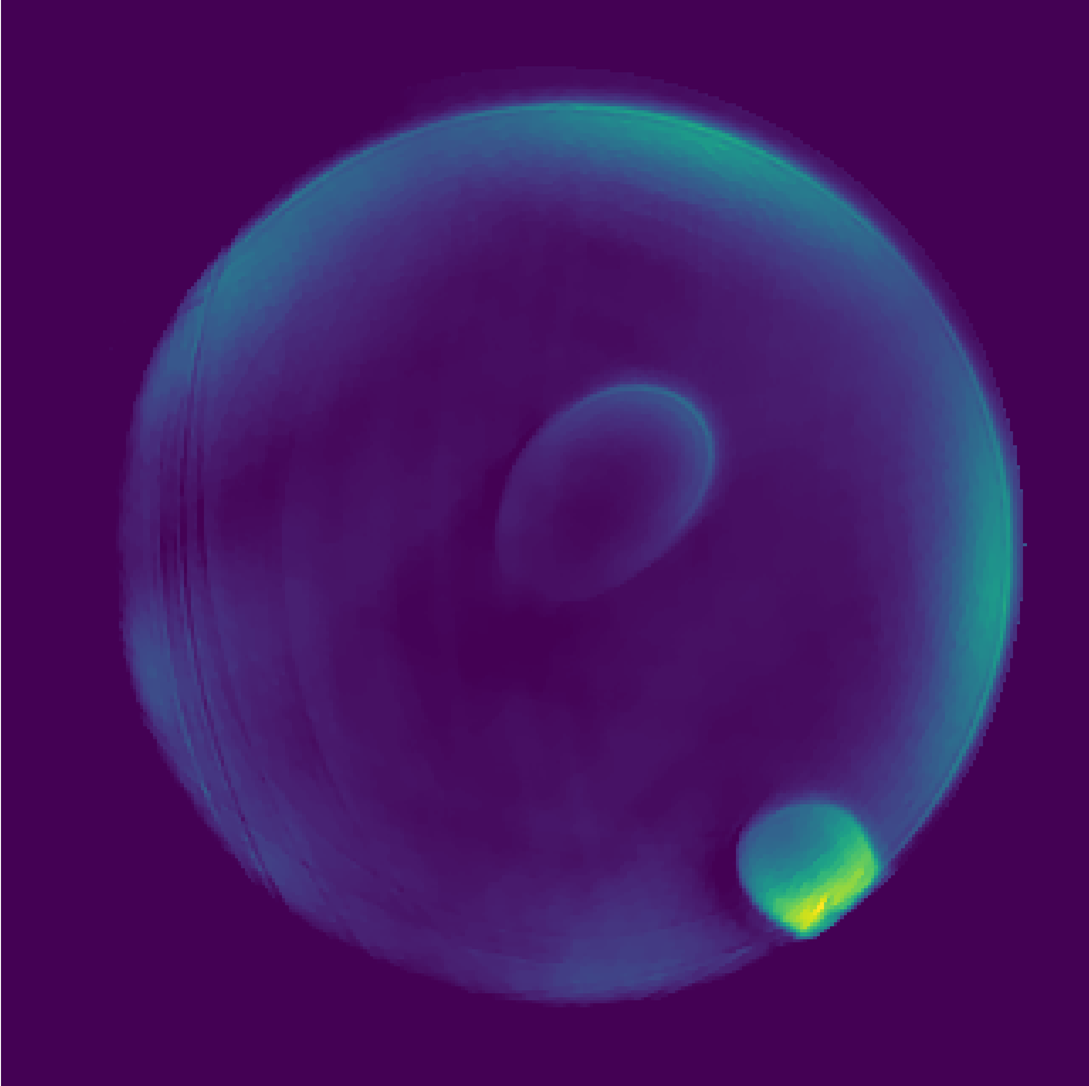}}\\

        \end{minipage}
        \begin{minipage}[c]{0.09\textwidth}
            \centering
            \hspace{0.15cm}\includegraphics[height=2.5cm]{images/scale_P.5.6.2_750.png}
        \end{minipage}
    \end{minipage}

    \caption{Comparison of reconstruction quality with 170 detectors out of the total 256 in use (179.3$^{\circ}$ angular coverage).}
    \label{fig:180_comparison}
\end{figure}

\begin{table}[h!]
    \footnotesize
    \centering
    \setlength{\tabcolsep}{10pt}   
    \renewcommand{\arraystretch}{1.5} 
    \setlength{\extrarowheight}{4pt}  

    \begin{tabular}{p{1.5cm} p{2.2cm}|ccc}
        \makecell{}
        & \makecell{\textbf{IQA}}
        & \makecell{\textbf{Initial}\\ \textbf{(FFT)} } 
        & \makecell{\textbf{DIP} }
        & \makecell{\textbf{TV} } \\
        \hline

        \makecell[l]{\textbf{Sample 1} } 
        & \makecell[l]{\rule{0pt}{3.2ex}$\uparrow$ PSNR (dB)  \\ $\uparrow$ SSIM\\ $\downarrow$ LPIPS \\ $\uparrow$ CC \\ $\uparrow$ HaarPSI } 
        & \makecell[l]{\rule{0pt}{3.2ex}13.373  \\ 0.587\\  0.378 \\  0.611\\ 0.175}
        & \makecell[l]{\rule{0pt}{3.2ex}\textbf{14.790} \\  \textbf{0.689}\\  \textbf{0.276} \\  0.623\\ \textbf{0.249}}
        & \makecell[l]{\rule{0pt}{3.2ex}14.396 \\ 0.652 \\0.281 \\\textbf{0.625} \\0.236}
        \\
        \hline

        \makecell[l]{\textbf{Sample 2} } 
        & \makecell[l]{\rule{0pt}{3.2ex}$\uparrow$ PSNR (dB)   \\ $\uparrow$ SSIM \\ $\downarrow$ LPIPS \\ $\uparrow$ CC \\ $\uparrow$ HaarPSI} 
        & \makecell[l]{\rule{0pt}{3.2ex}19.577  \\ 0.720 \\ 0.322 \\0.697 \\ 0.264}
        & \makecell[l]{\rule{0pt}{3.2ex}\textbf{22.063} \\ \textbf{0.812} \\ 0.230 \\ 0.710\\\textbf{0.320}}
        & \makecell[l]{\rule{0pt}{3.2ex}21.637 \\ 0.794 \\ \textbf{0.229} \\\textbf{0.715}\\0.314}

    \end{tabular}
    \caption{IQA measures (PSNR, SSIM, LPIPS, CC, HaarPSI) w.r.t.\ ground truth using DIP and TV regularization under 179.3$^{\circ}$ detector coverage. All IQA measures are computed over ROI. The arrows indicate the direction of improvement for each IQA metric, i.e., does a smaller or bigger value imply better image quality.}
    \label{table:180}
\end{table}

The results for the $179.3^{\circ}$ case are shown in Figure~\ref{fig:180_comparison} and Table~\ref{table:180}. DIP uses $\lambda=0.25$ for both samples 1 and 2. DIP was run to the maximum number of iterations (400). TV uses $\alpha=0.1$ for Sample 1 and $\alpha=0.25$ for Sample 2. Convergence required 367 iterations for Sample 1 and 282 iterations for Sample 2. All IQA measures are computed within the ROI.

For Sample 1, the initial reconstruction exhibits pronounced limited-view artifacts, most visible in streak-like distortions in the top-left region of the phantom. Both DIP and TV substantially reduce these artifacts, although some mild non-uniformity remains. In Sample 2, the initial reconstruction is comparatively clean and does not show obvious limited-view artifacts, but more loss of detail in the inner inclusion, and a dimmer appearance in general. Across both samples, DIP consistently delivers sharper boundaries for the inner inclusions compared to TV. While TV suppresses artifacts effectively, it also oversmooths fine structures, leading to visibly softer edges. DIP, on the other hand, preserves boundary sharpness better, which is reflected in its higher PSNR, SSIM, and HaarPSI scores. TV, on the other hand, is able to outperform DIP in CC for both samples and in LPIPS for Sample 2.

\subsubsection{112/256 detectors in use ($118.125^{\circ}$ angular coverage)}

\begin{figure}[h!]
    \centering

    \begin{minipage}[c]{0.08\textwidth}
        \centering
         \rotatebox{90}{\raisebox{0.0cm}{\scriptsize \textbf{Sample 1}}}
    \end{minipage}
    \begin{minipage}[c]{0.90\textwidth}
        \begin{minipage}[c]{0.2\textwidth}
            \centering
            \scriptsize \textbf{Ground truth \\ \phantom{}}\\
            \vspace{0.2em}
            \rotatebox{0}{
            \includegraphics[height=2.5cm, width=2.5cm, clip]{images/ground_truth_P.5.32_750.png}}\\
        \end{minipage}
        \hspace{0.002\textwidth}
        \begin{minipage}[c]{0.2\textwidth}
            \centering
            \scriptsize \textbf{Initial reconstruction}\\
            \vspace{0.2em}
            \rotatebox{0}{
            \includegraphics[height=2.5cm, width=2.5cm, clip]{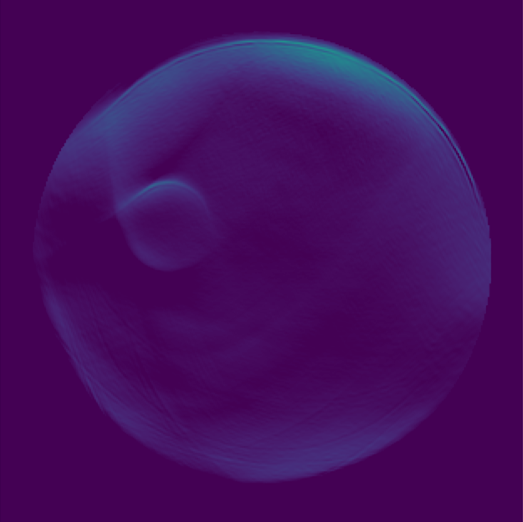}}\\

        \end{minipage}
        \hspace{0.002\textwidth}
        \begin{minipage}[c]{0.2\textwidth}
            \centering
            \scriptsize \textbf{DIP \\ \phantom{}}\\
            \vspace{0.2em}
            \rotatebox{0}{
            \includegraphics[height=2.5cm, width=2.5cm, clip]{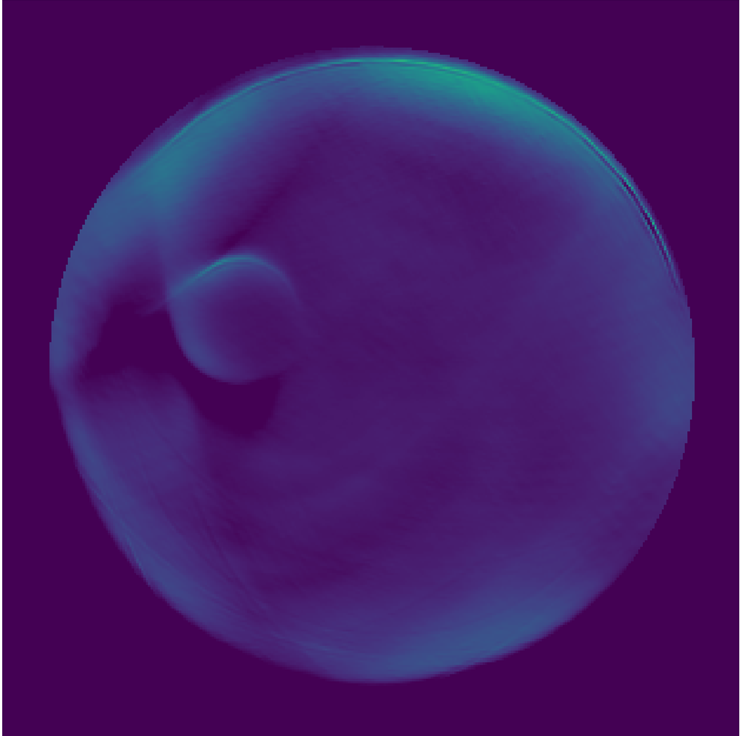}}\\

        \end{minipage}
        \hspace{0.002\textwidth}
        \begin{minipage}[c]{0.2\textwidth}
            \centering
            \scriptsize \textbf{TV \\ \phantom{}}\\
            \vspace{0.2em}
            \rotatebox{0}{
            \includegraphics[height=2.5cm, width=2.5cm, clip]{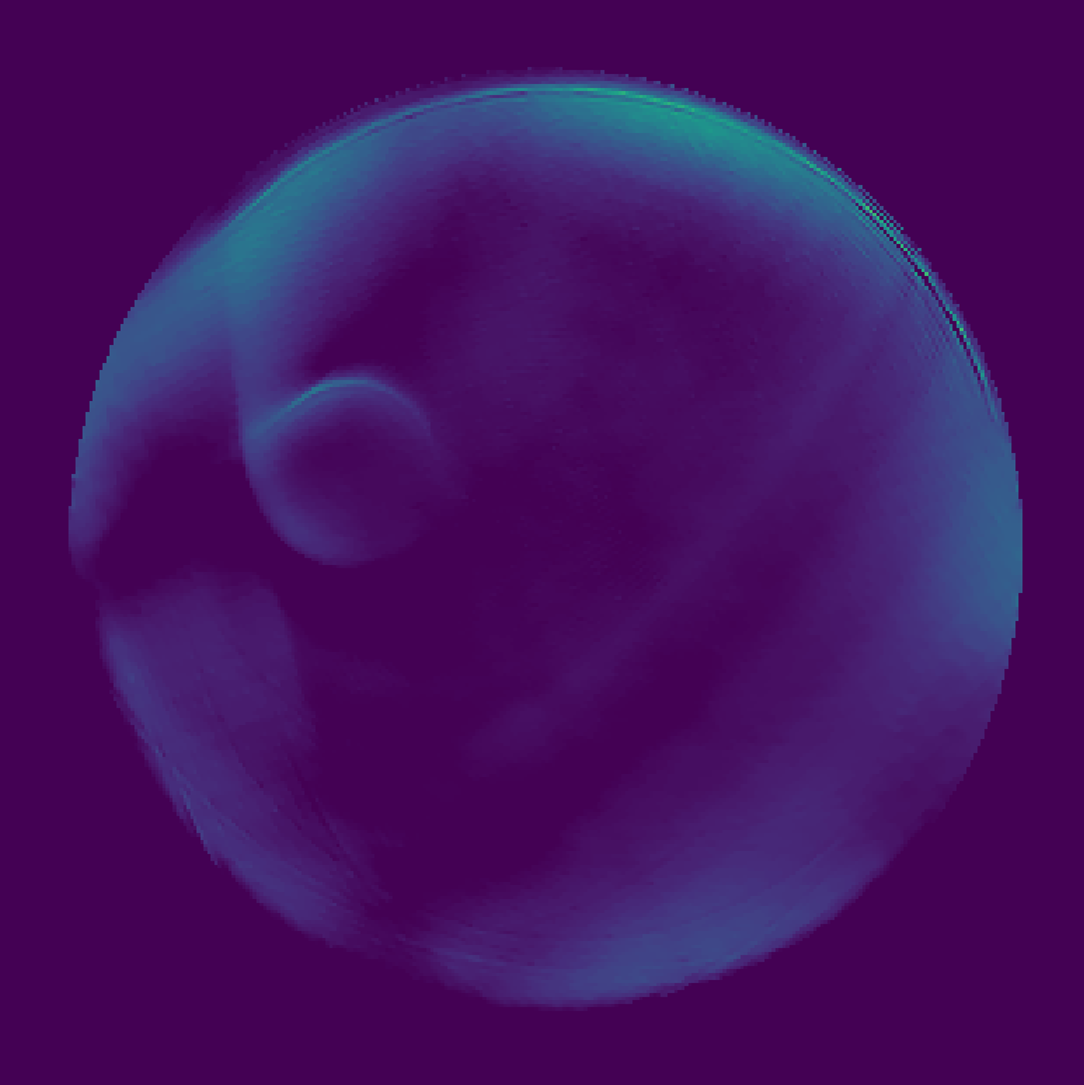}}\\
        \end{minipage}
        \begin{minipage}[c]{0.09\textwidth}
            \centering
            \vspace*{0.5cm}\includegraphics[height=2.5cm]{images/scale_P.5.32_750.png}
            
        \end{minipage}
    \end{minipage}

    \vspace{0.5em}

    \begin{minipage}[c]{0.08\textwidth}
        \centering
        \rotatebox{90}{\raisebox{0.0cm}{\scriptsize \textbf{Sample 2}}}
    \end{minipage}
    \begin{minipage}[c]{0.90\textwidth}
        \begin{minipage}[c]{0.2\textwidth}
            \centering
            \rotatebox{0}{
            \includegraphics[height=2.5cm, width=2.5cm, clip]{images/ground_truth_P.5.6.2_750.png}}\\
        \end{minipage}
        \hspace{0.002\textwidth}
        \begin{minipage}[c]{0.2\textwidth}
            \centering
            \rotatebox{0}{
            \includegraphics[height=2.5cm, width=2.5cm, clip]{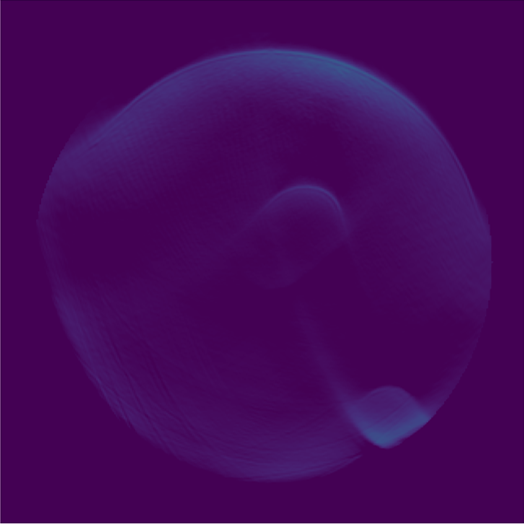}}\\

        \end{minipage}
        \hspace{0.002\textwidth}
        \begin{minipage}[c]{0.2\textwidth}
            \centering
            \rotatebox{0}{
            \includegraphics[height=2.5cm, width=2.5cm, clip]{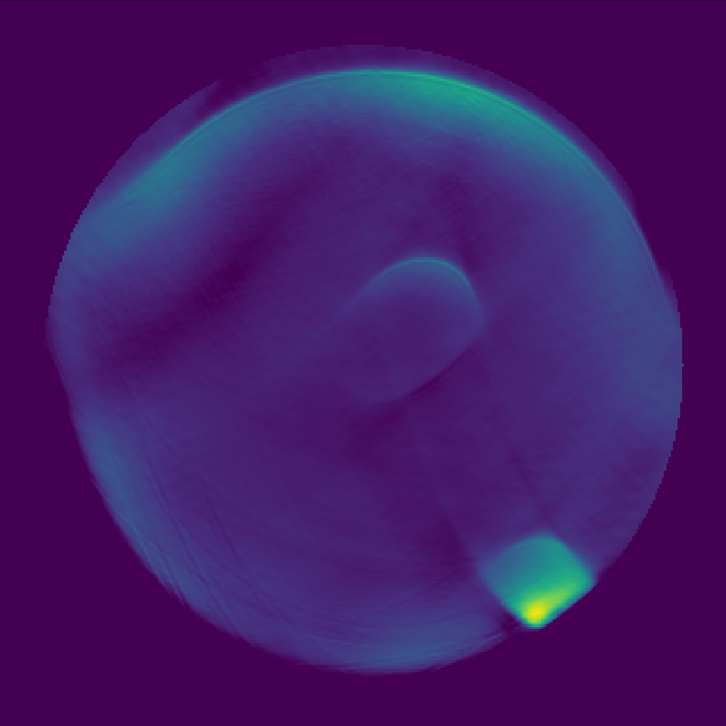}}\\

        \end{minipage}
        \hspace{0.002\textwidth}
        \begin{minipage}[c]{0.2\textwidth}
            \centering
            \rotatebox{0}{
            \includegraphics[height=2.5cm, width=2.5cm, clip]{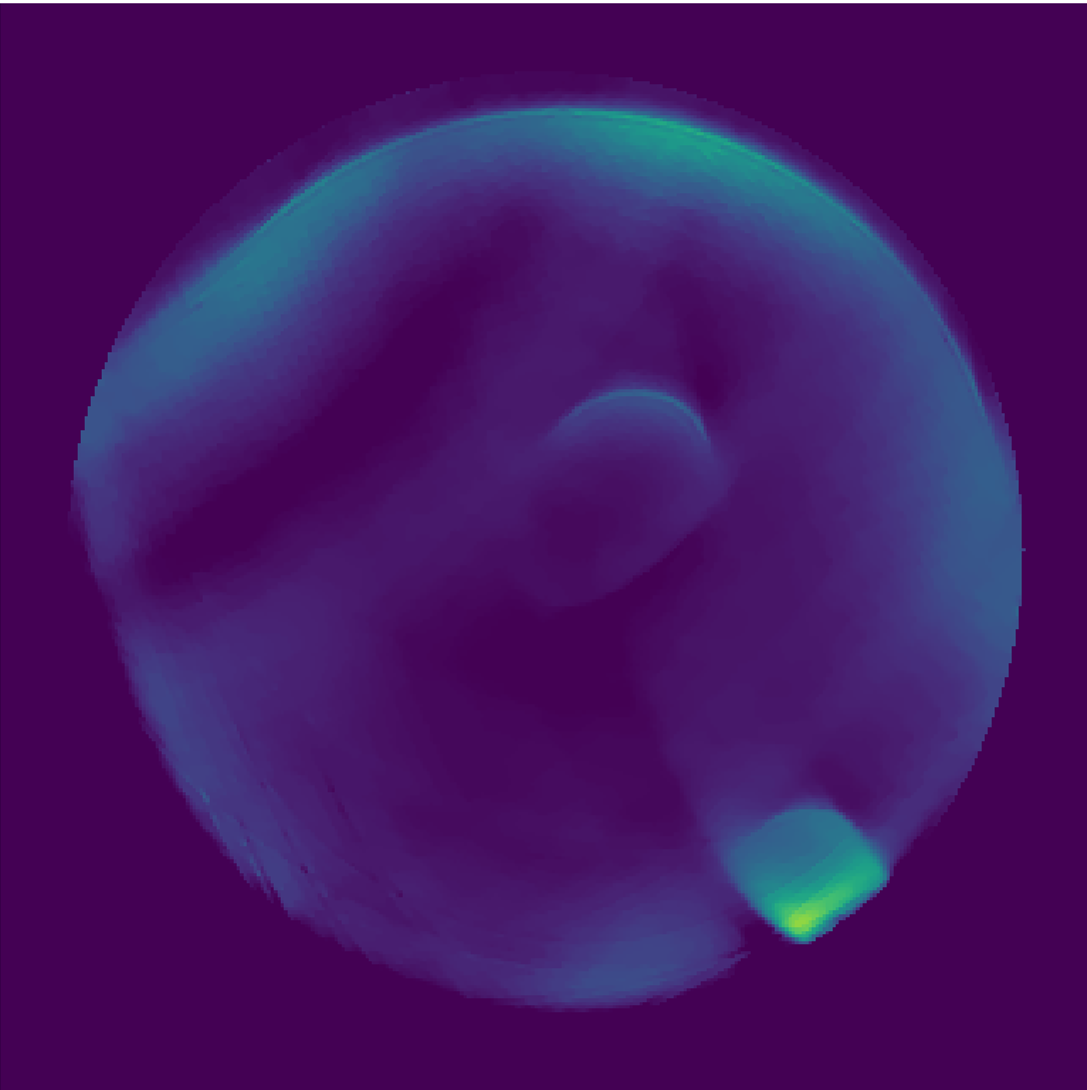}}\\
        \end{minipage}
        \begin{minipage}[c]{0.09\textwidth}
            \centering
            \hspace{0.15cm}\includegraphics[height=2.5cm]{images/scale_P.5.6.2_750.png}
        \end{minipage}
    \end{minipage}

    \caption{Comparison of reconstruction quality with 112 detectors out of the total 256 in use (118.125$^{\circ}$ angular coverage).}
    \label{fig:120_comparison}
\end{figure}

\begin{table}[h!]
    \footnotesize
    \centering
    \setlength{\tabcolsep}{10pt}   
    \renewcommand{\arraystretch}{1.5} 
    \setlength{\extrarowheight}{4pt}  

    \begin{tabular}{p{1.5cm} p{2.2cm}|ccc}
        \makecell{}
        & \makecell{\textbf{IQA}}
        & \makecell{\textbf{Initial}\\ \textbf{(FFT)} } 
        & \makecell{\textbf{DIP} }
        & \makecell{\textbf{TV} } \\
        \hline

        \makecell[l]{\textbf{Sample 1} } 
        & \makecell[l]{\rule{0pt}{3.2ex}$\uparrow$ PSNR (dB)  \\ $\uparrow$ SSIM\\ $\downarrow$ LPIPS \\ $\uparrow$ CC \\ $\uparrow$ HaarPSI } 
        & \makecell[l]{\rule{0pt}{3.2ex}12.962  \\ 0.538\\  0.445 \\  0.539\\ 0.139}
        & \makecell[l]{\rule{0pt}{3.2ex}\textbf{14.865} \\  \textbf{0.688}\\  \textbf{0.288} \\  \textbf{0.648}\\ \textbf{0.237}}
        & \makecell[l]{\rule{0pt}{3.2ex}14.367 \\ 0.634 \\0.301 \\0.642 \\0.223}
        \\
        \hline

        \makecell[l]{\textbf{Sample 2} } 
        & \makecell[l]{\rule{0pt}{3.2ex}$\uparrow$ PSNR (dB)   \\ $\uparrow$ SSIM \\ $\downarrow$ LPIPS \\ $\uparrow$ CC \\ $\uparrow$ HaarPSI} 
        & \makecell[l]{\rule{0pt}{3.2ex}18.641  \\ 0.646 \\ 0.390 \\0.604 \\ 0.215}
        & \makecell[l]{\rule{0pt}{3.2ex}\textbf{22.015}  \\ \textbf{0.799} \\ 0.255\\ 0.699\\\textbf{0.299}}
        & \makecell[l]{\rule{0pt}{3.2ex}21.471 \\ 0.782 \\ \textbf{0.253} \\\textbf{0.703 }\\0.298}

    \end{tabular}
    \caption{IQA measures (PSNR, SSIM, LPIPS, CC, HaarPSI) w.r.t.\ ground truth using DIP and TV regularization under 118.125$^{\circ}$ detector coverage. All IQA measures are computed over ROI. The arrows indicate the direction of improvement for each IQA metric, i.e., does a smaller or bigger value imply better image quality.}
    \label{table:120}
\end{table}

Finally, we show the results for the strongest limited view in Figure~\ref{fig:120_comparison} and Table~\ref{table:120}. For DIP, both samples use $\lambda=0.25$. TV uses $\alpha=0.1$ and 985 iterations for Sample 1, and $\alpha=0.25$ and 986 iterations for Sample 2. DIP was run to the maximum number of iterations (400). All IQA measures are computed within the ROI.

With only 112 detectors out of the total 256 in use, the reconstruction problem becomes severely ill-posed under strong limited-view conditions, and this is clearly reflected in the degradation of the initial reconstructions. In Sample 1, the initial reconstruction exhibits strong streak artifacts and loss of contrast, particularly around the central inclusion and boundary regions. DIP significantly improves the initial reconstruction by reducing these artifacts and restoring the overall circular shape, although some residual streaks remain. TV also reduces some of the limited-view artifacts, but the produced reconstruction lacks sharpness around the inclusion. This behavior is also reflected in the quantitative measures, all of which are in favor of DIP, although differences are subtle, especially in CC and HaarPSI.

In Sample 2, the limited-view geometry again leads to conspicuous artifacts in the initial reconstruction, but both DIP and TV successfully mitigate them, although some streak artifacts remain. Visually, DIP and TV produce quite similar reconstructions: TV reconstruction is slightly dimmer, but edge preservation is comparable between both methods. The similar visual performance yields competitive quantitative values, especially for LPIPS, CC, and HaarPSI.

\subsection{Experimental mouse data}

Figure \ref{fig: mouse data} compares reconstructions obtained using the initial FFT inverse reconstruction, DIP reconstruction with Leaky ReLU activation in the output layer, and DIP reconstruction without any nonlinearity in the output layer. A regularization parameter $\lambda=0.001$ was used in all cases. 

Across all scans, DIP with Leaky ReLU is able to reduce limited-view artifacts that are present in the initial reconstructions. Simultaneously, the main structural boundaries are retained, contrast is preserved, and the background is more homogeneous. Without nonlinearity, DIP is also able to improve on the initial reconstruction, although there are stronger dark bias and remaining limited-view streaks compared to the DIP reconstructions with Leaky ReLU. However, some fine details that get lost with nonlinear DIP, like the fine blood vessels in the kidneys in Scan 3, are still visible in the linear DIP reconstruction.

The TV reconstruction strongly suppresses artifacts in the background, but simultaneously, some important structural details are lost. Compared to DIP reconstructions, the visibility of smaller features is reduced and boundaries appear less sharp.

\begin{figure}[h!]
    \centering

    \begin{minipage}[c]{0.08\textwidth}
        \centering
         \rotatebox{90}{\raisebox{0.0cm}{\scriptsize \textbf{Scan 1}}}
    \end{minipage}
    \begin{minipage}[c]{0.90\textwidth}
        \begin{minipage}[c]{0.2\textwidth}
            \centering
            \scriptsize \textbf{Initial reconstruction \\ \phantom{}}\\
            \vspace{0.2em}
            \rotatebox{0}{
            \includegraphics[height=2.5cm, width=2.5cm, clip]{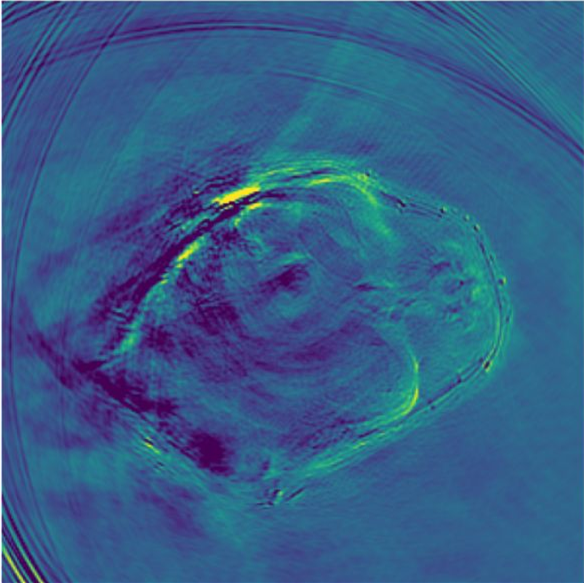}}\\
           
        \end{minipage}
        \hspace{0.002\textwidth}
        \begin{minipage}[c]{0.2\textwidth}
            \centering
            \scriptsize \textbf{DIP + Leaky ReLU \\ \phantom{}}\\
            \vspace{0.2em}
            \rotatebox{0}{
            \includegraphics[height=2.5cm, width=2.5cm, clip]{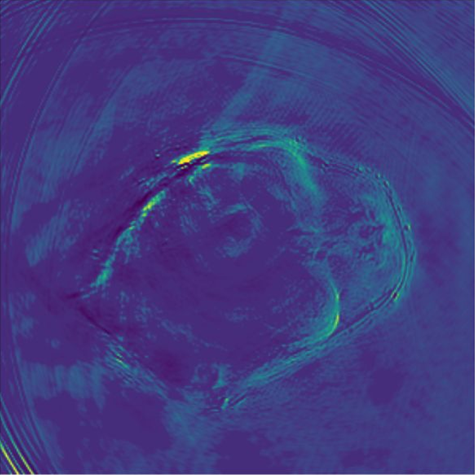}}\\

        \end{minipage}
        \hspace{0.002\textwidth}
        \begin{minipage}[c]{0.2\textwidth}
            \centering
            \scriptsize \textbf{DIP without nonlinearity \\ \phantom{}}\\
            \vspace{0.2em}
            \rotatebox{0}{
            \includegraphics[height=2.5cm, width=2.5cm, clip]{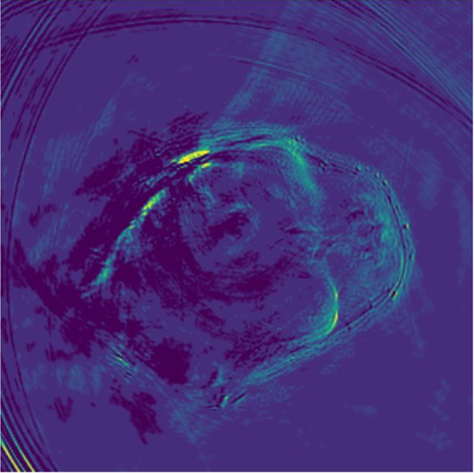}}\\

        \end{minipage}
        \hspace{0.002\textwidth}
        \begin{minipage}[c]{0.2\textwidth}
            \centering
            \scriptsize \textbf{TV \\ \phantom{} \\ \phantom{}}\\
            \vspace{0.2em}
            \rotatebox{0}{
            \includegraphics[height=2.5cm, width=2.5cm, clip]{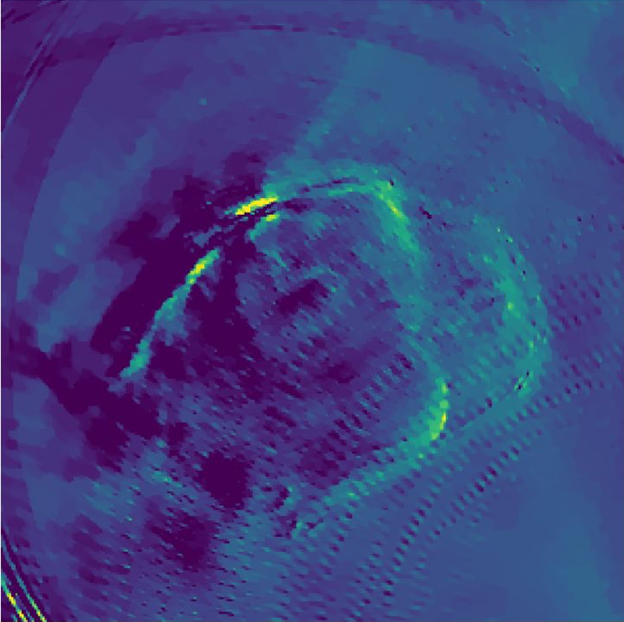}}\\
            
        \end{minipage}
        \begin{minipage}[c]{0.09\textwidth}
            \centering
            \vspace*{0.8cm}{\includegraphics[height=2.4cm]{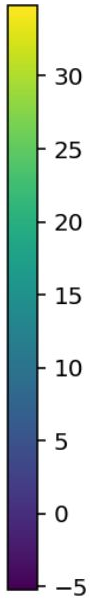}}

        \end{minipage}
    \end{minipage}

    \vspace{0.5em}

    \begin{minipage}[c]{0.08\textwidth}
        \centering
        \rotatebox{90}{\raisebox{0.0cm}{\scriptsize \textbf{Scan 2}}}
    \end{minipage}
    \begin{minipage}[c]{0.90\textwidth}
        \begin{minipage}[c]{0.2\textwidth}
            \centering
            \rotatebox{0}{
            \includegraphics[height=2.5cm, width=2.5cm, clip]{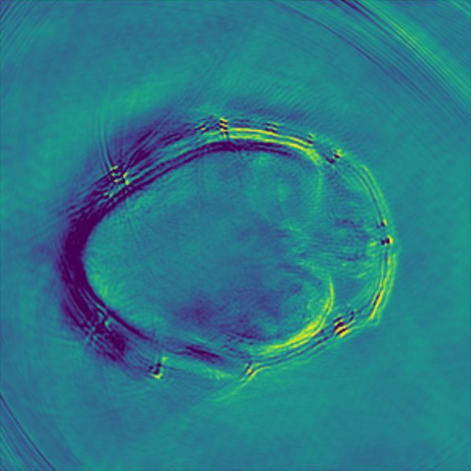}}\\
           
        \end{minipage}
        \hspace{0.002\textwidth}
        \begin{minipage}[c]{0.2\textwidth}
            \centering
            \rotatebox{0}{
            \includegraphics[height=2.5cm, width=2.5cm, clip]{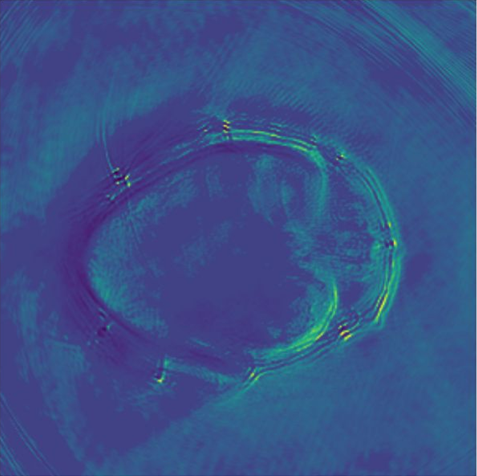}}\\

        \end{minipage}
        \hspace{0.002\textwidth}
        \begin{minipage}[c]{0.2\textwidth}
            \centering
            \rotatebox{0}{
            \includegraphics[height=2.5cm, width=2.5cm, clip]{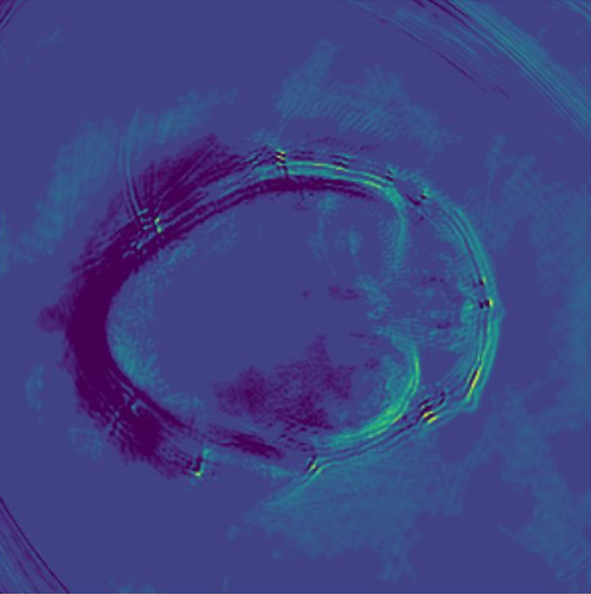}}\\

        \end{minipage}
        \hspace{0.002\textwidth}
        \begin{minipage}[c]{0.2\textwidth}
            \centering
            \rotatebox{0}{
            \includegraphics[height=2.5cm, width=2.5cm, clip]{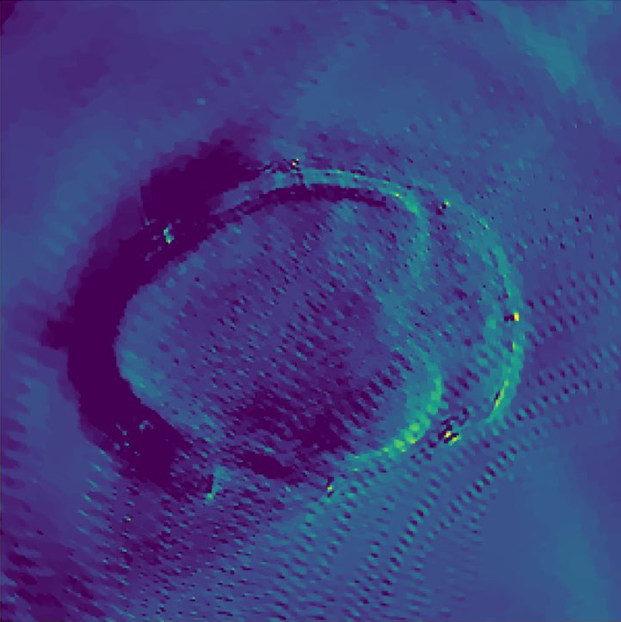}}\\
            
        \end{minipage}
        \begin{minipage}[c]{0.09\textwidth}
            \centering
            \hspace{0.1cm}\includegraphics[height=2.4cm] {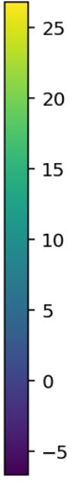}
          
        \end{minipage}
    \end{minipage}

    \vspace{0.5em}

    \begin{minipage}[c]{0.08\textwidth}
        \centering
        \rotatebox{90}{\raisebox{0.0cm}{\scriptsize \textbf{Scan 3}}}
    \end{minipage}
    \begin{minipage}[c]{0.90\textwidth}
        \begin{minipage}[c]{0.2\textwidth}
            \centering
            \rotatebox{0}{
            \includegraphics[height=2.5cm, width=2.5cm, clip]{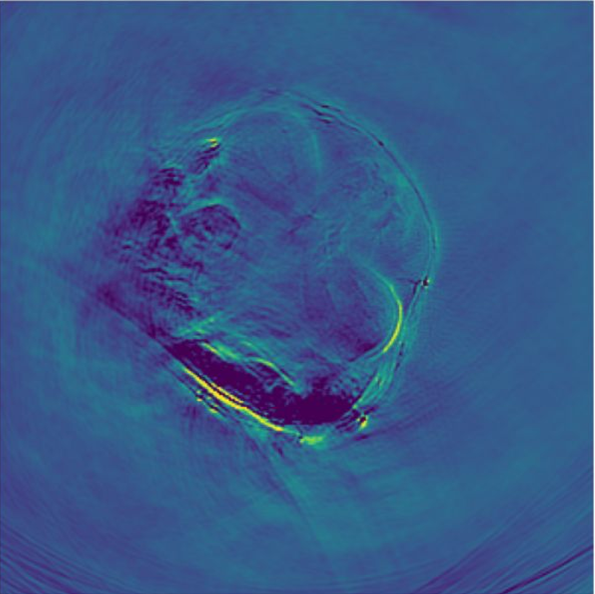}}\\
           
        \end{minipage}
        \hspace{0.002\textwidth}
        \begin{minipage}[c]{0.2\textwidth}
            \centering
            \rotatebox{0}{
            \includegraphics[height=2.5cm, width=2.5cm, clip]{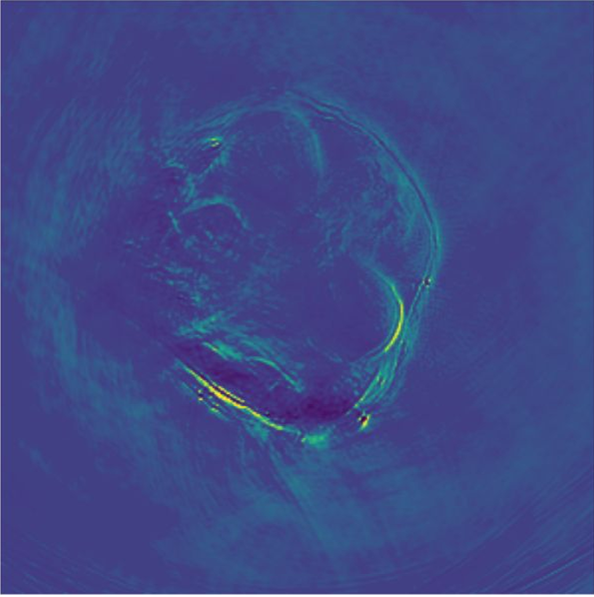}}\\

        \end{minipage}
        \hspace{0.002\textwidth}
        \begin{minipage}[c]{0.2\textwidth}
            \centering
            \rotatebox{0}{
            \includegraphics[height=2.5cm, width=2.5cm, clip]{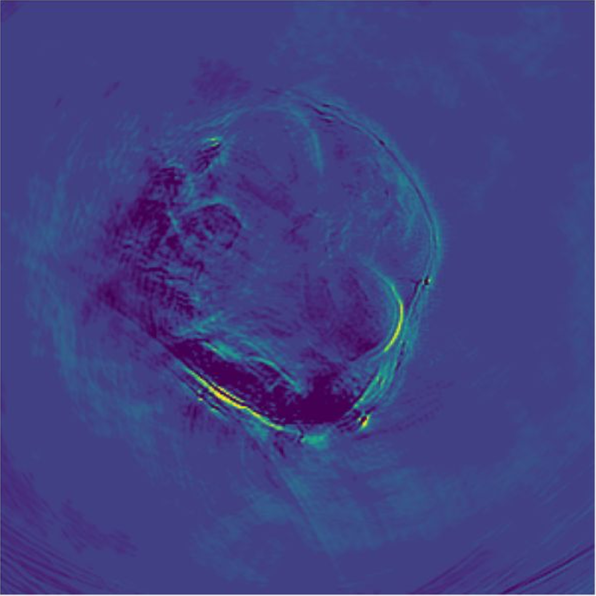}}\\

        \end{minipage}
        \hspace{0.002\textwidth}
        \begin{minipage}[c]{0.2\textwidth}
            \centering
            \rotatebox{0}{
            \includegraphics[height=2.5cm, width=2.5cm, clip]{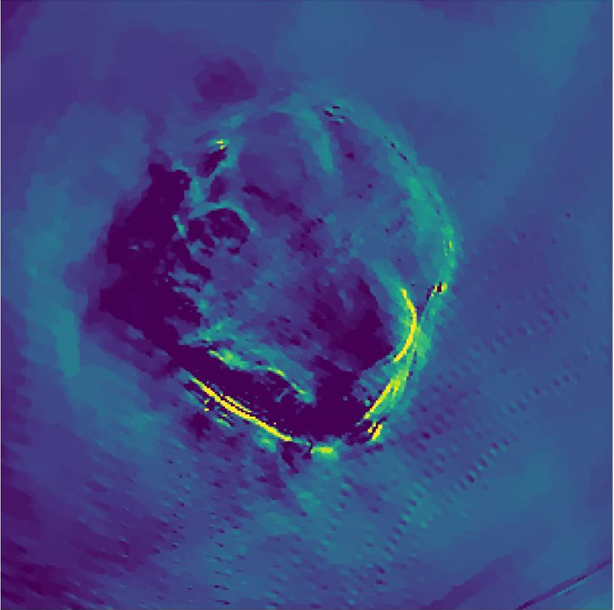}}\\
            
        \end{minipage}
        \begin{minipage}[c]{0.09\textwidth}
            \centering
            \hspace{0.15cm}\includegraphics[height=2.4cm]{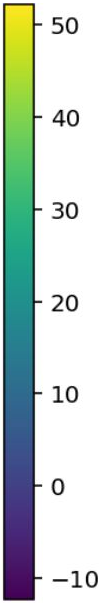}
          
        \end{minipage}
    \end{minipage}

\caption{Mouse data with 270$^{\circ}$ detector coverage comparing initial reconstruction with two DIP reconstructions: one with leaky ReLU activation function and one without any nonlinearity in the output layer. DIP used 600 iterations and $\lambda=0.001$ in all cases. TV uses $\alpha=2.0$ for all cases, and 2964, 3080, 2964 iterations for Scan 1, Scan 2 and Scan 3, respectively.}
\label{fig: mouse data}
\end{figure}

\FloatBarrier
\section{Discussion}

In this study, we compared two reconstruction strategies for noisy and limited-view PAT: a classic variational TV reconstruction and the Deep Image Prior (DIP) using an untrained convolutional neural network. Our results accentuate fundamental differences in their computational cost and reconstruction characteristics.

\subsection{Experimental data}
With experimental phantom data, TV regularization often reached its stopping criterion with fewer iterations compared to our maximum iteration approach with DIP. TV also exhibited predictable and stable optimization behavior. This is expected due to the convexity of the TV functional. As a result, TV reconstructions typically require only a relatively modest number of iterations and allow the use of reliable stopping rules free of ground truth. In contrast, DIP optimization is more unpredictable in parts due to the nonlinear neural network, which results in a non-convex optimization problem as well as the implicit bias of the underlying network architecture. In practice, this makes it difficult to define universal stopping criteria that do not depend on ground truth. In our experimental setting, we therefore adopted a fixed iteration limit of 400 iterations as a compromise: for some cases, the best reconstructions occurred well before the stopping criterion was met, while others continued improving beyond our cutoff.

The phantom experiments under 270$^{\circ}$ detector coverage and varying levels of additive noise (0\%, 10\%, and 20\%) revealed complementary strengths between the methods. DIP with early stopping consistently achieved the highest PSNR and SSIM values across all noise levels, which is consistent with the implicit bias of neural networks and their tendency to first fit low-frequency components and large-scale structures before overfitting to the noise. Early stopping thus acts as a form of implicit regularization that efficiently suppresses high-frequency noise while preserving the main structures of the phantom. Correlation coefficient (CC) remained similar across all methods and noise levels, indicating comparable global contrast. Despite excelling in PSNR and SSIM, DIP with early stopping falls behind both converged DIP and TV in perceptual metrics. 
This is consistent with the qualitative observation that early stopping with DIP produces images with slightly reduced local contrast, which translates to poorer performance in HaarPSI and LPIPS. Converged DIP and TV produced nearly identical perceptual and structural metrics. This similarity reflects the fact that both methods encourage piece-wise constant solutions through the TV regularization term. 

With the experimental phantom data, DIP consistently delivered superior reconstructions compared to TV, despite its higher computational cost. DIP was found to suppress limited-view artifacts more effectively while providing brighter contrast and better edge preservation. This behavior reflects the intrinsic bias of convolutional neural networks for preferring low-frequency solutions during optimization. TV, on the other hand, uniformly penalizes image gradients, which not only results in removing streak artifacts but also attenuates object edges and details. While \cite{breger2025study} highlights the challenges of individual IQA measures for applications such as PAT, the consistent advantage of DIP across several complementary metrics in our study points toward its superior performance. The use of multiple metrics together with qualitative assessment, therefore, provides a more robust and credible basis for our conclusion. Furthermore, as we investigated PSNR as a function of the regularization parameter in the range of 10$^{-3}$ to 10$^{0}$, we found no systematic changes. With all regularization parameter choices, DIP yielded a higher PSNR value compared to that of TV.

We also found that including an explicit TV regularizer inside the DIP loss widened the window for good reconstructions. This is most likely due to the fact that the TV regularizer balances DIP's tendency to overfit high-frequency components - including noise - at the later stages of the optimization process, meaning that reconstructions remain stable across a much larger range of iterations. Therefore, the need for early stopping is relaxed, and good reconstructions can be obtained without excessive tuning of the stopping point. 

\subsection{\textit{In vivo} measurements}

The \textit{in vivo} mouse experiments presented additional challenges compared to the phantom experiments. These include fine anatomical structures to be resolved (e.g., renal vascular features) and the presence of strong acoustic reflections. The reflection artifacts are caused, for example, by trapped air in the intestines and in the ultrasound gel inside the clingfilm used for acoustic coupling. The reflections introduce severe reconstruction artifacts, particularly on the stomach side of the mouse, leading to negative reconstructed pressures and completely non-recoverable contrast in parts of the reconstruction. 

Despite these data limitations, DIP was able to recover several anatomically relevant structures, especially towards the back of the mouse, where signal quality was higher. However, fine vessel-like structures that are still visible in the initial reconstruction were either attenuated or lost during the process. There are several possible factors for this. First, the mouse abdomen only occupies a small portion of the full, uncropped reconstruction domain, and the background contains artifact-dominated areas. The DIP could be enticed to focus on mitigating artifacts in the background, which might compromise reconstruction quality in the mouse abdomen. Second, the fine vessels are small and structurally similar to the limited-view artifacts that DIP architecture is inherently targeting. As small vessels and artifacts are both high-frequency components that the network only starts to learn during overfitting, it may be forced to compromise between artifact suppression and preservation of fine detail. Discarding the nonlinearity in the output layer of our neural network did help retain some of the finest vessel structures, as visible in Figure \ref{fig: mouse data}, but this comes at the expense of artifact mitigation.

As expected, the classical TV was also able to suppress noise and artifacts in the reconstructions. However, edges and fine structures were attenuated even more strongly than with DIP reconstructions, leading to overly smooth reconstructions. Additionally, with the experimental mouse data, TV required substantially more iterations to meet the convergence criterion.

\subsection{Model optimization}

The DIP model used in this study employed a single off-the-shelf U-Net architecture with a generic set of hyperparameters not optimized for this task. It is therefore possible that a different network and set of parameters could produce slightly different results. 

Finally, runtime measurements confirm that DIP requires a higher computational budget. The runtime of both DIP and primal-dual TV reconstruction algorithms with experimental phantom data was evaluated over 1000 iterations. Reconstruction times were measured on NVIDIA RTX A5000 GPU using PyTorch 2.5.1 with CUDA 11.8. Computation of a thousand iterations required 142.9 seconds for DIP and 130.5 seconds for primal-dual.

\section{Conclusions}

We demonstrate that the Deep Image Prior (DIP) framework can be successfully applied to circular-geometry photoacoustic tomography (PAT) with the recently published fast forward and adjoint algorithms, even under challenging limited-view setups. Across our experiments, the DIP framework consistently reduced limited-view artifacts and produced reconstructions with improved ground truth fidelity, brightness, and edge preservation compared to the classical variational TV-regularized primal-dual reconstruction, especially so for experimental phantom data. These qualitative observations are supported by our set of quantitative indicators, all pointing towards better performance of DIP over TV in the experimental phantom data case and most of the simulation studies.

The results underline a clear trade-off. Classical TV regularization offers a fast, predictable, and robust baseline: it converges quickly and predictably compared to DIP, and thus allows the use of straightforward stopping criteria that do not require access to the ground truth. On the contrary, DIP requires more iterations, but in most cases provides stronger suppression of limited-view streaking and often better preservation of fine structures and edges. Importantly, adding an explicit TV-regularization term inside the DIP objective relaxes the need for early stopping, and a plain maximum iteration cutoff works well enough for all experimental data cases. In the \textit{in vivo} mouse experiments, reconstruction quality was limited by factors like severe reflection artifacts and trapped air, resulting in non-recoverable contrast in some regions. Nevertheless, DIP demonstrated robust artifact suppression and the ability to simultaneously recover relevant anatomical structures in areas where the measurements were less severely corrupted by noise.

While our experiments demonstrate promising reconstruction performance for DIP, definite statements about stability and broader applicability require additional work. Further studies are needed to form a comprehensive understanding of the convergence properties, generalizability, and possibly a more suitable stopping criterion. Our findings indicate that DIP is a promising unsupervised reconstruction strategy for experimental PAT, even for limited-view acquisition geometries. 
\section*{Acknowledgments}
This work has been supported in part by the
Research Council of Finland (Project Nos. 353093, 353086,  Finnish Centre of Excellence in Inverse Modelling and Imaging; and Project Nos. 359186, 358944, Flagship of Advanced Mathematics for Sensing Imaging and Modelling; Project No. 338408, Academy Research Fellow (AI-SOL)), the European Research Council (ERC) under the European Union’s Horizon 2020 Research and Innovation Programme (Grant Agreement No. 101001417—QUANTOM), and the Finnish Ministry of Education and Culture’s Pilot for Doctoral Programmes (Pilot project Mathematics of Sensing, Imaging and Modelling). 
LK is partially supported by the NSF through the Award No. NSF/DMS-2405348. The work of JG was supported by the Deutsche Forschungsgemeinschaft, Germany (DFG, German Research Foundation) under projects GR 5824/1 and GR 5824/2.

\medskip
Received xxxx 20xx; revised xxxx 20xx; early access xxxx 20xx.
\medskip

\end{document}